\documentclass{article}
\usepackage{amsmath,amssymb,amsfonts}
\usepackage{arxiv}
\usepackage{algorithm}
\usepackage{algorithmic}
\usepackage[utf8]{inputenc} % allow utf-8 input
\usepackage[T1]{fontenc}    % use 8-bit T1 fonts
\usepackage{hyperref}       % hyperlinks
\usepackage{url}            % simple URL typesetting
\usepackage{booktabs}       % professional-quality tables       % blackboard math symbols
\usepackage{nicefrac}       % compact symbols for 1/2, etc.
\usepackage{microtype}      % microtypography
\usepackage{lipsum}		% Can be removed after putting your text content
\usepackage[numbers]{natbib}
\usepackage{doi}
\usepackage{graphicx}   % For including graphics
\usepackage{caption}    % For customizing captions
\usepackage{cleveref}   % For improved cross-referencing
\usepackage{array}
\usepackage{tabularx}
\usepackage{adjustbox}
\usepackage{enumitem}
\usepackage{longtable}
\usepackage{geometry}
\usepackage{booktabs} % For professional looking tables
\usepackage{tabularx} % For flexible column widths
\usepackage{caption} % For better table captions
\usepackage{CJK}
\usepackage{cleveref}

\newcommand{\figref}[1]{\hyperref[#1]{Figure\ref*{#1}}}

\newcommand{\tabref}[1]{\hyperref[#1]{Table\ref*{#1}}}

\newcommand{\Algref}[1]{\hyperref[#1]{Algorithm\ref*{#1}}}
\renewcommand{\arraystretch}{1.3} % Adjust the row height
\title{Optimizing Urban Mobility Through Complex Network Analysis and Big Data from Smart Cards}

\author{ \href{https://orcid.org/}{\includegraphics[scale=0.06]{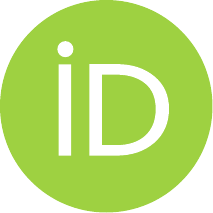}\hspace{1mm}Li Sun}, 
\href{https://orcid.org/0009-0003-8414-2996}{\includegraphics[scale=0.06]{orcid.pdf}\hspace{1mm}Negin Ashrafi}, 
\href{https://orcid.org/0009-0003-7159-3245}{\includegraphics[scale=0.06]{orcid.pdf}\hspace{1mm}Maryam Pishgar} \\
Department of Industrial and Systems Engineering \\
University of Southern California, Los Angeles, CA 90089 \\
\texttt{\{lsun4765, ashrafin, pishgar\}@usc.edu} \\
}
\renewcommand{\undertitle}{Research Paper}
\begin{document}
\maketitle

\begin{abstract}
This study investigates the network characteristics of high-frequency (HF) and low-frequency (LF) travelers in urban public transport systems through the analysis of 20 million smart card records from Beijing’s transit network. A novel methodology integrates advanced data preprocessing, clustering techniques, and complex network analysis to differentiate HF and LF passenger behaviors and their impacts on network structure, robustness, and efficiency. The primary challenge lies in accurately segmenting and modeling the behaviors of diverse passenger groups within a large-scale, noisy dataset while maintaining computational efficiency and scalability. HF networks, representing the top 25\% of travelers by usage frequency, exhibit high connectivity (average clustering coefficient: 0.72) and greater node degree centrality but lower robustness, with efficiency declining by 35\% under targeted disruptions and longer average path lengths (6.2) during peak hours. Conversely, LF networks, comprising 75\% of travelers, are more dispersed yet resilient, with efficiency declining by only 10\% under similar disruptions and stronger intracommunity connectivity. Temporal analysis reveals that HF passengers significantly contribute to peak-hour congestion, with 57.4\% of HF trips occurring between 6:00--10:00 AM, while LF passengers exhibit broader temporal dispersion, mitigating congestion hotspots. Understanding these travel patterns is crucial for optimizing public transit systems. The findings suggest targeted strategies, such as enhancing robustness in HF networks by diversifying key routes and improving accessibility in LF-dominated areas. This research provides a scalable framework for analyzing smart card data and offers actionable insights into urban transit system design, congestion management, and sustainable mobility planning.
\end{abstract}


\maketitle

\section{Introduction}

In the contemporary framework of urban transportation strategy, the prioritization of public transportation development has become a key approach to mitigating urban traffic congestion. Elevating the role of public transit within the urban transportation mix is now essential, underpinned by national policies that advocate for the expansion of public transportation systems focused on meeting passenger needs \cite{b1}. As cities experience increases in population density and a diversification of travel destinations, the limitations of traditional single-mode bus networks become increasingly evident, prompting a shift towards more scalable, multimodal public transportation networks. This evolution has facilitated the emergence of medium-to-high capacity transit solutions such as Bus Rapid Transit (BRT) and trams, leading to the development of integrated multimodal networks characterized by distinct operational entities, methods, and functional positioning. Today, dual-mode public transit systems are prevalent across many urban centers, reflecting a broader trend towards spatial dispersion and growth in the scale of urban activities. Public transport networks, conceptualized as complex systems, have been extensively analyzed in both conventional bus networks \cite{b2} and rail transit networks \cite{b3}, providing critical insights into the dynamics of urban mobility and the structural evolution of public transport infrastructure.

In urban public transport, the adoption of smart card technology has significantly enhanced travel convenience by enabling precise recording of card swiping times and locations. These systems, often integrated with GPS and transit management systems, improve operational efficiency and resource management. Despite these advancements, gaps remain, particularly in capturing alighting stations due to the prevalent flat-fare system \cite{b4}. Delving deeper into smart card data reveals detailed travel patterns and distribution of demand, providing insights into the distinct impacts of different traveler frequencies on the network's efficiency, connectivity, and traffic distribution. Recent studies highlight how evolving work patterns also influence transit demand \cite{b4!}. This analysis is crucial for optimizing the public transportation network to better cater to diverse user needs.

Smart card data, captured by Automated Fare Collection (AFC) systems, includes essential information such as card number, transaction date, and travel details like time, route, and station \cite{b4'}. Additionally, bus operation data from Automated Vehicle Location (AVL) systems offers comprehensive data on station locations and route specifics, enhancing the granularity of transit analyses \cite{b4''}. This research explores the distinctions between high and low-frequency travelers to better understand their impact on urban public transportation systems. High-frequency travelers often prioritize service regularity and punctuality, especially during peak hours, whereas low-frequency travelers may prefer routes that cater to comfort and accessibility, such as those leading to tourist destinations. Recognizing these preferences enables more tailored service delivery, which can enhance user satisfaction and system efficiency. Similarly, understanding behavioral factors influencing transportation choices is essential, as studies have shown that subjective attitudes significantly impact mobility decisions and policy acceptance \cite{b4'''}. By understanding these varied travel patterns, transport planners can implement more effective strategies, including route adjustments and schedule optimizations, to boost the public transport system’s appeal and competitive edge.

A comprehensive literature review underscores significant advancements in the field of big data processing and analysis of bus travel characteristics. Researchers have explored various facets of public transportation data to enhance the understanding and operational efficiency of urban transit systems.

Xuewu Chen from Southeast University has been pivotal, utilizing bus smart card data to develop methodologies that analyze passenger flows and travel information, thus setting a foundation for future research \cite{b5}. Building on this, Daixiao, also from Southeast University, delved into smart card data for individual bus routes, proposing innovative ways to capture key indicators of passenger flow \cite{b6, b7}. Further enriching this area, Liyun Zhu introduced GIS-based data mining technology to refine the analysis of bus passenger flow statistics and travel characteristics \cite{b8}.

In contrast, the Florida Department of Transportation’s ADAMS system illustrates an international approach, integrating data from AVL, APC, and AFC systems to enhance state-wide bus system operations, though it lacked a comprehensive study across different data sources \cite{b9}. Notably, researchers like Barry and Jinhua Zhao have merged smart card data with vehicle GPS information to create detailed models of travel trajectories and rail transit OD matrices, showcasing the potential of combining various data sources to improve transit system analysis \cite{b10, b11}.
Klein's work \cite{b12} in "Sensor Technologies and Data Requirements for ITS" introduced advanced data fusion techniques like Bayesian inference, neural networks, and fuzzy logic for intelligent transportation systems

The scope of research also extends to examining the dynamic and spatiotemporal characteristics of passenger travel behavior. Studies by Tuan Sheng Chen and others have categorized travel patterns into simple and complex chains, exploring the relationship between personal travel choices and transportation modes \cite{b13}. Chenchen Deng and Fei Xiao combined smart card and GPS data to visually analyze bus passenger flows, enhancing the understanding of how these flows vary across time and space \cite{b14}. Researchers like Ya Wen and Yingrong Dong have further developed models to predict passenger flow dynamics at transport hubs and analyzed non-work travel frequencies, respectively, using advanced statistical methods \cite{b18, b19}.

Moreover, the research by Atizaz Ali and Viallard focuses on understanding the stability of travel behaviors and the impact of various factors such as holidays on passenger travel frequency, employing clustering techniques like K-means to categorize travel patterns \cite{b20, b21}. Le Minh Kieu’s work using the DBSCAN algorithm to explore similarities in travel behaviors among passengers highlights the utility of smart card data in segmenting user groups based on their travel habits \cite{b22}. Menno Yap \cite{b23} and Howard Wong studied public transportation elasticity, highlighting the varying impacts of service changes on demand.

Gao Ziyou and Wu Jianjun \cite{b24} analyzed Beijing's public transportation network using complex network theory, showing scale-free characteristics in routes and small-world properties in stations. Bo Wang extended this by modeling transit systems in Beijing, Shanghai, and Hangzhou, emphasizing scale-free networks with preferential linking. Hui Li Yan \cite{b25} applied complex network theory to slow traffic networks in Xi'an, confirming small-world and scale-free characteristics and expanding its application in urban transportation. Similarly, Derrible and Kennedy \cite{b2} highlighted scale-free and small-world patterns in metro systems, emphasizing robust structures and transfer hubs.

Miao Y., Zhiru Wang, and Albert P.C. \cite{b27} proposed a vulnerability analysis framework for multi-modal networks based on accessibility, addressing disruptions and socio-economic impacts. Da Silva \cite{b28} modeled transit systems as complex networks, identifying congestion and inefficiencies in Curitiba’s network. Yap and Cats \cite{b29} utilized machine learning to predict disruptions and classified station vulnerabilities, providing actionable insights for improving transit operations.

Overall, the existing body of research demonstrates a robust exploration of bus smart card data and its applications, providing new perspectives on passenger flow analysis, travel trajectory estimation, and public transit system optimization. This collective effort underscores the critical role of advanced data processing and analysis techniques in enhancing the operational efficiency and service quality of urban public transportation systems.

Analyzing the existing scholarly research reveals significant gaps in understanding urban public transportation networks, specifically in two critical areas:

Firstly, there is an insufficient focus on the characteristics of high and low-frequency travelers. Much of the current research centers around empirical analysis and the construction of network models that primarily examine the physical structure's vulnerability through static and dynamic analyses. However, the nuances of high and low-frequency traveler behaviors are often overlooked due to challenges in maintaining a unique and consistent card identifier over time, which is not always available from public transportation operators. This limitation curtails in-depth investigations into the distinct behavioral patterns and impacts of different traveler groups within urban transit systems, underscoring the need for more detailed studies.

Secondly, the complex network characteristics of urban public transportation are not fully understood, particularly the multi-level and dynamic aspects of these networks. While many studies have focused on the static properties of network structures, such as nodes and connections, there is a lack of comprehensive research on the dynamic behaviors of networks, the patterns of passenger flow, and temporal shifts within these systems. The existing literature typically does not address the dynamic changes of complex networks over time and space, indicating a significant gap in current research methodologies.

These shortcomings emphasize the necessity for more robust data collection efforts, the application of advanced analysis techniques, and the development of more precise models to enhance the understanding of urban public transportation networks in future research.

In conclusion, despite advancements in urban public transportation research, crucial gaps remain, particularly in understanding traveler behavior and network dynamics. This research aims to bridge these gaps by employing advanced analytical techniques and improving data collection, thus enhancing our understanding of transit networks and improving urban transportation systems.

\section{Methodology}
Step-1: Data Preprocessing:Extract data from smart card records, including bus and subway travel information. Process the data by removing blanks and duplicate travel chains, ensuring each passenger's travel chain is complete and accurate

Step-2: Station Clustering:Apply the K-means algorithm to cluster all bus and subway stations in Beijing based on their latitude and longitude coordinates.

Step-3: Travel Frequency Classification: Classify passengers into high-frequency and low-frequency groups using smart card data, based on their travel frequency.

Step-4: Complex Network Model Construction: Build a network model to represent high- and low-frequency passenger travel patterns, capturing spatial and temporal network characteristics.

Step-5: Network Visualization: Visualize the network to display travel patterns and traffic flows, identifying key trends and relationships.

Step-6: Network Characteristics Analysis: Conduct a comprehensive analysis of network features, including:
6.1 Node charatertics analysis: A comprehensive analysis of node degree, betweenness centrality, and closeness centrality provides critical insights into the structural and operational characteristics of the transportation network. This approach identifies key nodes, which play a strategic role in maintaining network connectivity and managing flow, serving as critical hubs. The identification of such nodes lays a solid foundation for the subsequent robustness tests.
6.2 Robustness Test: Simulate network disruptions by removing key nodes and analyzing the impact on network stability.Ba
6.3 Basic Property Analysis: Evaluate fundamental network properties such as degree distribution, clustering coefficient, and path length.
6.4 Community Detection: Use the Louvain algorithm to identify community structures and interaction patterns.
6.5 Peak Hour Analysis: Examine network behavior during peak travel periods to understand traffic dynamics and passenger flow patterns.
This structured approach enables a detailed understanding of public transit networks and passenger behavior, as illustrated in Figure~\ref{fig:1}:

\begin{figure}[ht]
    \centering
    \includegraphics[width=15cm, height=10cm]{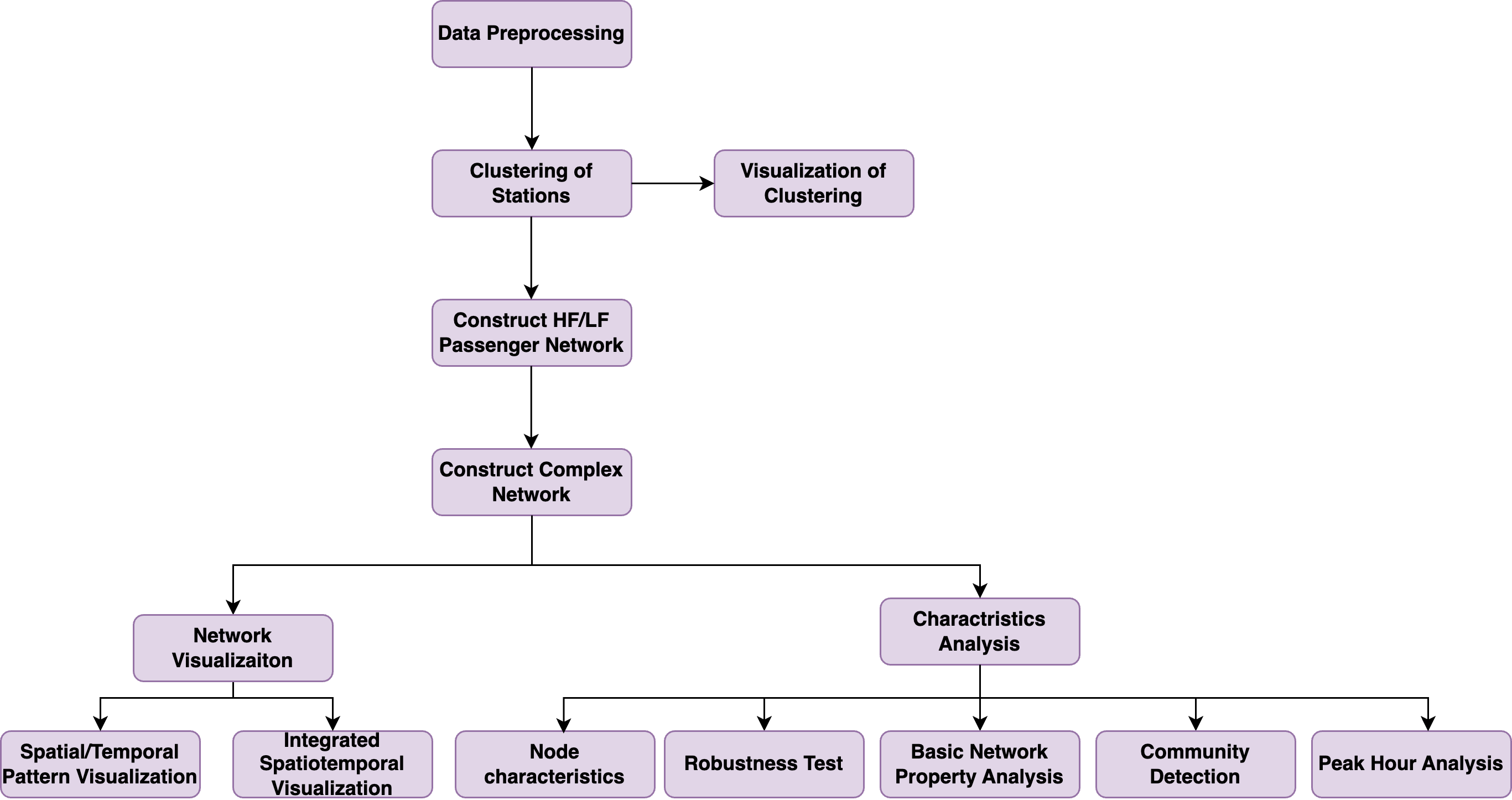}
    \caption{Flowchart of data preprocessing, clustering, and network analysis processes.}
    \label{fig:1} 
\end{figure}

\section{Research Data Pre-processing}
\subsection{Introduction to public transportation data}
The primary sources of public transport big data, including smart card data and station data from buses and subways, are often distributed across different systems with varied data formats. Given the diversity of these data sources and the inconsistency in formats, data preprocessing becomes crucial in ensuring the effectiveness of the research. This study employs database software to comprehensively preprocess public transportation big data, encompassing data cleaning, submission, and transformation.

This paper focuses on the Beijing public transport system as a case study, particularly examining the individual travel modes of passengers. It concentrates on the complete travel chain information of passengers over a two-week period in March 2018. A detailed examination of the structure and functionality of Beijing's intelligent bus system is presented, systematically describing the data collection process, screening criteria, and extraction methods used for the research datasets. This groundwork not only provides a robust data basis for identifying and analyzing travel patterns within the public transport system but also offers empirical support for studying passenger travel behavior and its impact on the transport network.

Building on the foundational research by Cui Meili \cite{b31}, the collection of public transportation data is categorized into static and dynamic components. Static data encompasses foundational information about bus and subway operations, including station coordinates, while dynamic data captures real-time elements such as GPS positioning of buses and subways, IC card swiping events, and vehicle speed. The meticulous collection and preprocessing of these data types are crucial for establishing a robust public transportation data analysis platform. This platform underpins comprehensive studies of the transportation system and supports the development of optimization strategies.

In related work by Ma Yiqing \cite{b32}, smart IC card data emerges as a critical source, offering detailed insights into passenger interactions within the transport system. This data includes variables such as IC card numbers, bus lines used, times of card usage, types of transactions, and expenditure amounts. A deep dive into these details enables a nuanced understanding of passenger travel behaviors. For instance, Table~\ref{tab:1} provides a thorough explanation of each field in the IC card dataset. In focusing on the primary objectives of this study, data elements that do not directly impact the research outcomes, such as expenditure amount, remaining balance, and transaction type, are excluded from the data extraction process. This selective approach to data handling sharpens the research focus and enhances the efficiency and accuracy of data processing.

\begin{table}[h]
\centering
\caption{IC Card Data Segments}
\label{tab:1}
\begin{tabular}{>{\centering\arraybackslash}p{3cm} >{\centering\arraybackslash}p{3cm} >{\centering\arraybackslash}p{6cm}}
\toprule
\textbf{Field Name} & \textbf{Data Type} & \textbf{Explanation} \\
\midrule
BUSNO         & Varchar2 & Vehicle Code \\
CARDNO        & Varchar2 & Card Number \\
CONSUME       & Number   & Amount Spent \\
CONSUMEDATE   & Date     & Transaction Time \\
CONSUMETYPE   & Number   & Transaction Type \\
LINENO        & Varchar2 & Line Code \\
REMAINTIMES   & Number   & Remaining Times \\
\bottomrule
\end{tabular}
\end{table}

\subsection{Research Data}
In this study, the data structure is meticulously designed to comprehensively capture and represent the travel characteristics of passengers within the transit system. As illustrated in Table~\ref{tab:2}, the data structure meticulously logs the travel activities of passengers. This includes an encrypted passenger identifier, the bus or subway line number, the sequence number and name of the boarding station on that specific line, and the precise time of card swiping, enriching the temporal dimension of the passengers' travel patterns.

For passengers engaging in transfers, the structure goes further to document the exact time of the transfer and details of the subsequent travel chain. This includes the sequence number of the transfer station and the time of transfer, thereby compiling a complete record of the passenger's travel route.

It is crucial to acknowledge the immense volume of data handled in this analysis—approximately 18 million rows of passenger travel records daily over a span of 14 days, culminating in about 20 million rows of data in total. This underscores the complexity and critical nature of subsequent data processing efforts needed to distill meaningful insights from such a vast dataset.

\begin{table}[h]
\small % Reduce font size
\setlength{\tabcolsep}{3pt} % Reduce column spacing for a tighter table
\centering
\caption{Passengers Data}
\label{tab:2}
\begin{tabularx}{\textwidth}{>{\raggedright\arraybackslash}p{5cm} 
                           >{\centering\arraybackslash}p{1cm}  
                           >{\centering\arraybackslash}p{1cm}  
                           >{\centering\arraybackslash}p{2cm} 
                           >{\raggedright\arraybackslash}p{2cm} 
                           >{\centering\arraybackslash}p{2cm} 
                           >{\centering\arraybackslash}p{2cm}}
\toprule
\textbf{Passengers ID} & \textbf{Mode} & \textbf{Line Num} & \textbf{Station Num} & \textbf{Station Name} & \textbf{Time} & \textbf{Next Travel Chain} \\
\midrule
\texttt{770fdeffe6154df9a2f631027035} & DT & 6   & 49 & Jintai Road            & 20180301090700 & DT \\
\texttt{770fdeffe6154df9a2f631027035} & DT & 14  & 67 & Dawang Road           & 20180301143200 & DT \\
\texttt{770fdeffe6154df9a2f631027035} & GJ & 423 & 10 & Xinfadi Bridge West   & 20180301164301 & GJ \\
\texttt{770fdeffe6154df9a2f631027035} & GJ & 22  & 1  & Mudanyuan             & 20180301185301 & GJ \\
\texttt{7ef24c978d18195279f53c5a0b3}  & DT & 6   & 57 & Huangqu               & 20180301123900 & DT \\
\texttt{770fdeffe6154df9a2f631027035} & DT & 6   & 49 & Jintai Road           & 20180301090700 & DT \\
\bottomrule
\end{tabularx}
\end{table}

As presented in Table~\ref{tab:3} and Table~\ref{tab:4}, the longitude and latitude of the stations provide precise locational data within the urban geographical space. This spatial information enables a detailed analysis of traffic flow between stations and serves as a foundation for exploring complex network characteristics, including connectivity, clustering coefficients, and shortest paths. By incorporating these coordinates into subsequent research, this study enhances the understanding of bus and subway network operation patterns from a physical space perspective, which is instrumental for effective public transportation planning and management.

Moreover, the availability of geographical coordinate data facilitates the creation of visual representations. Using MATLAB or similar visualization tools, station flow diagrams can be developed to illustrate the usage of individual stations and the movement patterns of passengers between different stations across the city. These visualizations offer valuable insights into spatial and temporal travel behaviors, aiding in identifying key transit hubs and optimizing the overall transportation network.

\begin{table}[h]
\centering
\caption{Subway Station Coordinates}
\label{tab:3}
\begin{tabularx}{\textwidth}{>{\raggedright\arraybackslash}X 
                           >{\centering\arraybackslash}X 
                           >{\centering\arraybackslash}X}
\toprule
\textbf{Station Name} & \textbf{Longitude} & \textbf{Latitude} \\
\midrule
Anhe Bridge North     & 116.269956 & 40.012195 \\
Beiyuan               & 116.277647 & 40.002373 \\
Guoyuan               & 116.290908 & 39.998258 \\
Yuanlingyuan          & 116.310186 & 39.999662 \\
Peking University East Gate & 116.315842 & 39.992212 \\
Shuguang Road         & 116.316467 & 39.983991 \\
Haidian Huangzhuang   & 116.317564 & 39.975996 \\
\bottomrule
\end{tabularx}
\end{table}

\begin{table}[h]
\centering
\caption{Bus Station Coordinates}
\label{tab:4}
\begin{tabularx}{\textwidth}{>{\raggedright\arraybackslash}X 
                           >{\centering\arraybackslash}X 
                           >{\centering\arraybackslash}X}
\toprule
\textbf{Station Name} & \textbf{Longitude} & \textbf{Latitude} \\
\midrule
Tiantongyuan          & 116.3465167 & 40.00291006 \\
Tiantongyuan Bridge West & 116.338111 & 39.89619858 \\
Xiaoying              & 116.328866 & 39.89615096 \\
Huanggang Village South Entrance & 116.5236192 & 39.92578004 \\
Dongbei Wang Middle Road & 116.2855868 & 40.03777604 \\
Hongfu Industry Garden & 116.500131 & 39.801502 \\
Liangxiang Bus Station & 116.527203 & 39.92557497 \\
\bottomrule
\end{tabularx}
\end{table}

\begin{figure}[ht]
    \centering
    \includegraphics[width=11cm, height=6.5cm]{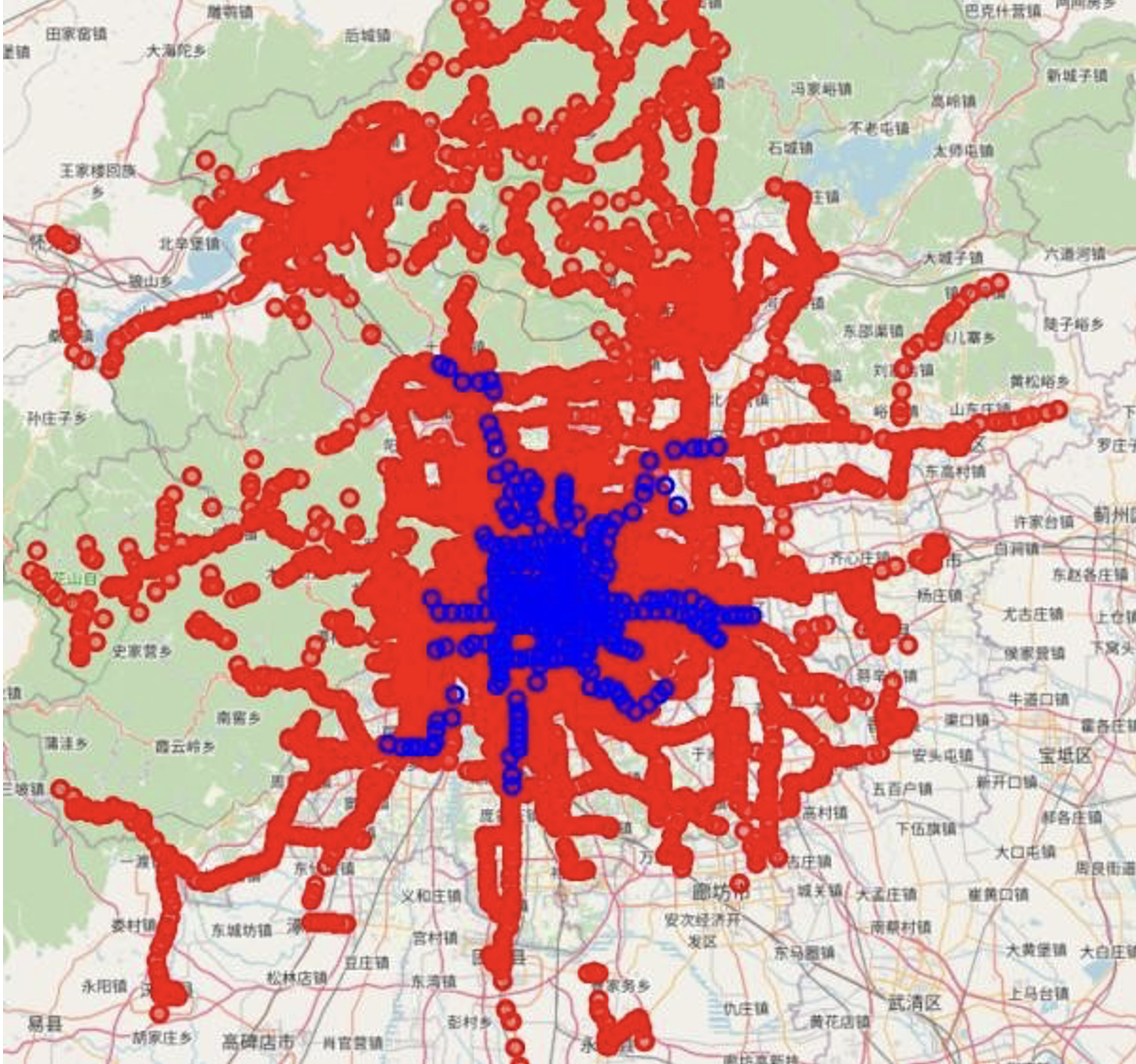}
    \caption{A GIS diagram of bus stops and subway stations.}
    \label{fig:2}
\end{figure}

The station information within the public transport system is visualized using geographic information system (GIS) technology, which accurately plots each subway and bus station on a map based on their latitude and longitude coordinates. This process generates an interactive spatial distribution map, where bus stops are initially represented in red and subway stations in blue. However, to ensure visual consistency and fluidity in subsequent analyses, all stations are uniformly displayed in green when traffic data is incorporated. This choice reflects the study's focus on overall network dynamics rather than distinguishing between subway and bus stations. The visualization results, as shown in Figure~\ref{fig:2}, provide a comprehensive spatial representation of station distribution, forming the basis for deeper exploration of public transportation patterns.

\subsection{Passenger Data Preprocessing}
In this study, data preprocessing encompasses several critical steps to ensure the reliability and consistency of the dataset, particularly for passenger travel records and the geographic data of bus and subway stations.

To begin with, fully duplicate records are identified and removed from the dataset. These duplicates often arise from errors during data acquisition, such as multiple recordings of the same data during upload. Eliminating these redundancies not only reduces the volume of unnecessary data but also enhances the accuracy and efficiency of subsequent analyses.

Additionally, data records with evident flaws are excluded. These flaws may stem from technical failures, user misoperations, or processing errors and include anomalies such as unrealistic timestamps, illogical station sequences, and implausible speed or distance calculations. Since such flawed data does not accurately represent actual travel conditions, its inclusion could mislead the analysis and compromise the validity of the findings.

For passenger travel data, incomplete travel chains are also discarded. Travel chains may appear incomplete due to various reasons, including technical interruptions or missing records during the data collection process. As these incomplete records cannot provide a full representation of passenger behavior, they are excluded from the study to maintain the integrity of the analysis.

By implementing these preprocessing steps, the study establishes a cleaner, consistent, and reliable dataset, forming a robust foundation for detailed traffic analysis and the examination of complex network properties.

\subsection{Stations Cluster}

Due to the complexity of passenger and site information, direct visualization often yields suboptimal results. To address this challenge, a clustering algorithm is applied to process site information, enabling the extraction of more meaningful patterns. The motivation for employing K-means clustering in this study lies in its ability to reduce data complexity and enhance computational and analytical efficiency. By aggregating a large number of individual sites into a smaller number of representative cluster nodes, the analysis not only becomes more manageable but also facilitates the identification and understanding of key nodes and connections within the urban transportation network.

In this study, the K-means clustering algorithm is employed due to its widespread use and effectiveness. The algorithm operates by dividing the dataset into K clusters, minimizing the sum of distances from each data point to its nearest centroid. The process begins with the random selection of K data points as initial centroids. Through iterative refinement, each data point is assigned to the nearest centroid cluster, followed by an update of the centroid positions based on the newly assigned clusters. This iterative process continues until convergence conditions are met, either when centroid updates no longer significantly change the clusters or when a preset number of iterations is reached. This approach ensures a systematic reduction of complexity while retaining critical insights into the network's structure and dynamics.

In this study, a mixed clustering approach was applied to 53,700 subway and bus stations, ultimately grouping them into 537 nodes. This method effectively simplifies the network model while preserving critical traffic flow information, making it feasible to conduct more focused and efficient network analysis and optimization. By reducing the complexity of the data, this clustering process ensures that the essential structural and operational characteristics of the transportation network are retained.

The results of the clustering process highlight the key nodes in the network, as partially shown in Table~\ref{tab:5}, with the detailed stations belonging to the first cluster center presented in Table~\ref{tab:6}. It is important to note that bus and subway stations were clustered together in this analysis. Despite their differences, this mixed clustering approach does not compromise the accuracy or effectiveness of the clustering results, demonstrating its robustness and suitability for analyzing integrated urban transportation networks.

% Table 2-5: The central node after clustering
\begin{table}[h]
\centering
\caption{The Central Node After Clustering}
\label{tab:5}
\begin{tabularx}{\textwidth}{>{\centering\arraybackslash}X 
                           >{\centering\arraybackslash}X 
                           >{\centering\arraybackslash}X}
\toprule
\textbf{Clustering Nodes} & \textbf{Latitude} & \textbf{Longitude} \\
\midrule
0 & 40.04867448 & 116.4066763 \\
1 & 40.44167793 & 115.9061828 \\
2 & 39.854609   & 116.3362191 \\
3 & 40.35755938 & 116.5678062 \\
4 & 39.97114007 & 115.9879883 \\
5 & 39.87181985 & 116.6642087 \\
\bottomrule
\end{tabularx}
\end{table}

% Table 2-6: The first cluster center internal site
\begin{table}[h]
\centering
\caption{The First Cluster Center Internal Site}
\label{tab:6}
\begin{tabularx}{\textwidth}{>{\centering\arraybackslash}X 
                           >{\raggedright\arraybackslash}X 
                           >{\centering\arraybackslash}X 
                           >{\centering\arraybackslash}X 
                           >{\centering\arraybackslash}X}
\toprule
\textbf{Stop Number} & \textbf{Stop Name} & \textbf{Longitude} & \textbf{Latitude} & \textbf{Node Number} \\
\midrule
42388 & Xisan Village      & 116.5894841 & 40.36254838 & 0 \\
42399 & Xisan Village      & 116.5898399 & 40.36246982 & 0 \\
42400 & Koutou             & 116.570183  & 40.36211652 & 0 \\
42547 & Guanhe Crossing    & 116.554231  & 40.36048595 & 0 \\
42548 & Sanhe Crossing     & 116.5489953 & 40.37759114 & 0 \\
42581 & Beizhai Village East & 116.5588375 & 40.336021   & 0 \\
\bottomrule
\end{tabularx}
\end{table}

During the clustering process, the new centroid of each node is determined by calculating the mean of all sites belonging to the cluster. This includes averaging the longitude and latitude of the sites and summing their respective traffic flows. Consequently, the position of each node represents the geometric center of all sites within the cluster, while the node's traffic reflects the cumulative flow of these sites. Additionally, the inter-station traffic data is standardized using a maximum-minimum normalization approach after clustering.

This approach ensures that the longitude, latitude, and flow data of each node not only capture the spatial distribution of traffic but also provide insights into the degree of traffic aggregation within the network. By combining spatial and flow characteristics, this method enhances the understanding of network dynamics and facilitates the analysis of inter-node interactions. The clustering results, including inter-node flow data, are summarized in Table~\ref{tab:7}.

\begin{table}[h]
\centering
\small
\setlength{\tabcolsep}{3pt} % 减少列间距
\caption{Inter-node Flow Rate After Clustering}
\label{tab:7}
\begin{tabularx}{\textwidth}{>{\centering\arraybackslash}p{1cm} 
                             >{\centering\arraybackslash}p{1.5cm} 
                             >{\centering\arraybackslash}p{2cm} 
                             >{\centering\arraybackslash}p{2cm} 
                             >{\centering\arraybackslash}p{2cm} 
                             >{\centering\arraybackslash}p{2cm} 
                             >{\centering\arraybackslash}p{1.5cm} 
                             >{\centering\arraybackslash}p{2.5cm}} % 扩展 Standardized Traffic
\toprule
\textbf{Start-site Node ID} & \textbf{Terminate-site Node ID} & \textbf{Start-site Longitude} & \textbf{Start-site Latitude} & \textbf{Terminate-site Longitude} & \textbf{Terminate-site Latitude} & \textbf{Flow} & \textbf{Standardized Traffic} \\
\midrule
121 & 234 & 116.453518 & 39.9081348 & 116.470691 & 39.91255031 & 56811 & 1 \\
234 & 121 & 116.470691 & 39.91255031 & 116.453518 & 39.9081348 & 51013 & 0.897940503 \\
121 & 42 & 116.453518 & 39.9081348 & 116.630711 & 39.89759435 & 37126 & 0.653494103 \\
234 & 144 & 116.470691 & 39.91255031 & 116.7877687 & 39.9473497 & 33524 & 0.5908098773 \\
234 & 470 & 116.470691 & 39.91255031 & 116.498658 & 39.91021268 & 31525 & 0.5549023063 \\
245 & 234 & 116.4698259 & 39.93250931 & 116.470691 & 39.91255031 & 29423 & 0.517910778 \\
245 & 234 & 116.470691 & 39.91255031 & 116.4698259 & 39.93250931 & 29378 & 0.517910664 \\
470 & 234 & 116.498658 & 39.91021268 & 116.470691 & 39.91255031 & 28940 & 0.509397513 \\
121 & 7 & 116.453518 & 39.9081348 & 116.4569124 & 39.89013758 & 28234 & 0.496972364 \\
7 & 121 & 116.4569124 & 39.89013758 & 116.453518 & 39.9081348 & 28160 & 0.495669776 \\
\bottomrule
\end{tabularx}
\label{table:flow_rate}
\end{table}

This study undertakes a comprehensive preprocessing of intelligent IC card data for buses and subways, urban road network data, and the location information of bus and subway stations in Beijing for March 2018. The preprocessing steps include removing duplicate records, correcting incomplete travel chains, and excluding records with obvious defects, thereby ensuring the quality and accuracy of the data for subsequent analysis. Additionally, the structure of the dataset is carefully outlined, encompassing fields such as passenger ID, transportation mode or route, station number and name, travel time, and subsequent travel chain details. These fields provide essential information for analyzing passenger travel patterns.

\subsection{Section Summary}
The dataset comprises approximately 1,800,000 daily records spanning 14 days, totaling nearly 200,000,000 records, offering a robust basis for this research. Latitude and longitude information for 288 subway stations and 53,443 bus stations in Beijing is collected and processed, providing a physical spatial foundation for analyzing inter-station traffic and complex network characteristics. This spatial data also enables the visualization of traffic flows. To further simplify the network model, the K-means clustering algorithm is employed to cluster 53,700 subway and bus stations into 537 nodes. This method effectively reduces data complexity while preserving critical traffic flow information, facilitating detailed network analysis and optimization.

The preprocessing efforts establish a strong data foundation, supporting the analysis of high- and low-frequency travel modes, complex network characteristics, and the formulation of service optimization strategies in subsequent analyses.

\section{Network Construction}
\subsection{High and low frequency passenger networks}

This study successfully constructs a daily travel chain information database for bus passengers through in-depth mining and analysis of big data from the public transportation system. The database utilizes a unified storage format, organizing the data based on passengers' IC card numbers, thereby forming a structured and extensive database of bus travel chain information. Within this system, it becomes straightforward to retrieve the complete travel trajectory of any passenger on a specific date and perform detailed analyses of the spatiotemporal distribution characteristics of transit travel.

In addition to the creation of this extensive bus travel chain database, the study also emphasizes the development of high and low-frequency passenger networks. By distinguishing passengers based on their travel frequency, two distinct sub-networks are formed: one for high-frequency passengers who use public transportation regularly, and another for low-frequency passengers who infrequently rely on the bus system. This differentiation facilitates a more detailed examination of passenger travel behaviors and provides targeted data to support the precise planning and optimization of the bus network. The aim of constructing these high and low-frequency passenger networks is to better understand the travel patterns and needs of different passenger groups. This insight is crucial for developing more efficient, personalized operational strategies, enhancing service quality, and contributing to the sustainable development of urban public transportation systems.

\begin{figure}[h]
    \centering
    \includegraphics[width=10cm, height=9cm]{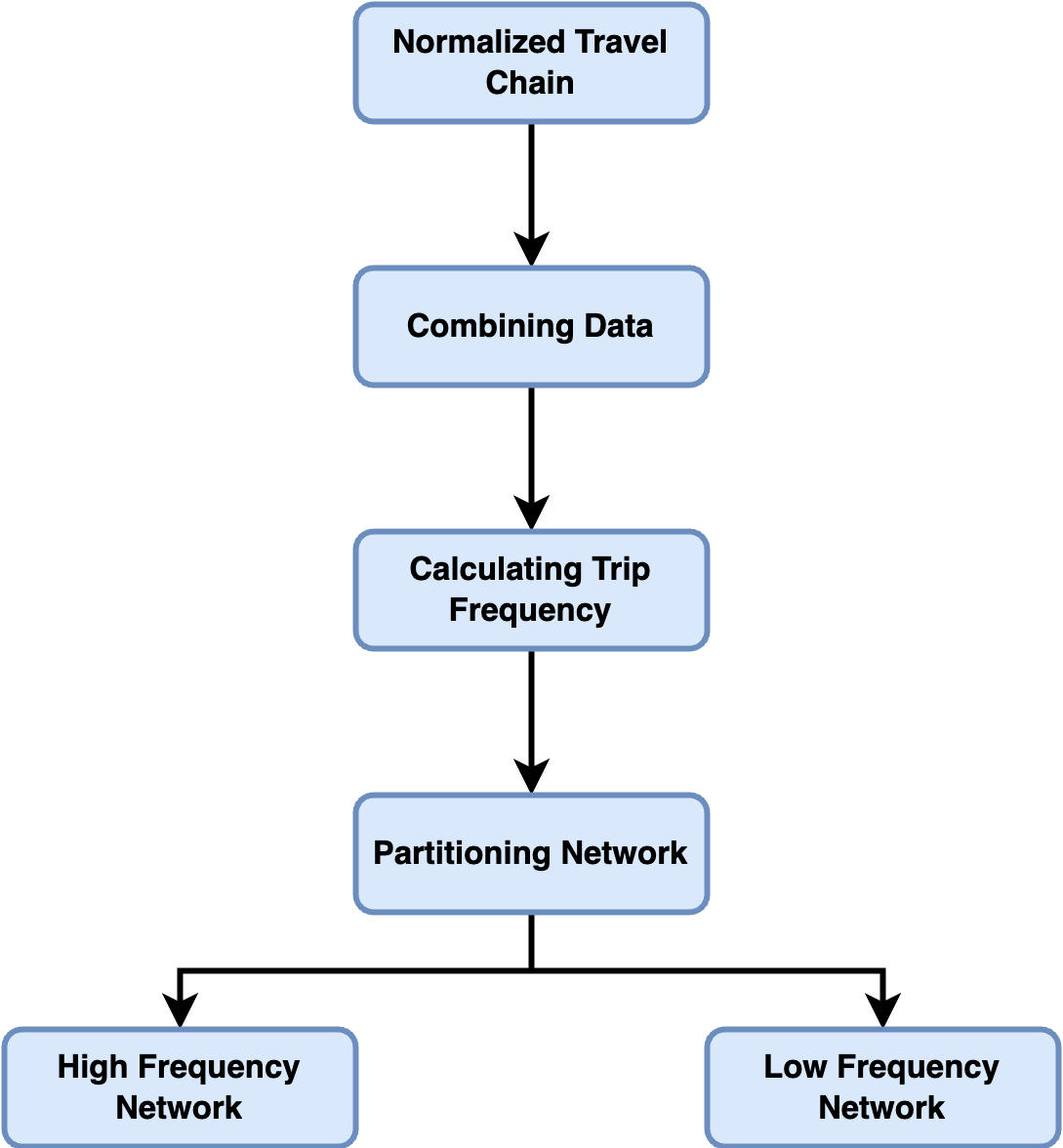}
    \caption{A GIS diagram of bus stops and subway stations.}
    \label{fig:3}
\end{figure}

 Each step was designed to improve the accuracy and effectiveness of the data analysis, ensuring a robust and reliable dataset for subsequent analysis. These steps are comprehensively illustrated in Figure~\ref{fig:3}, which provides a visual representation of the detailed process.

Step 1 - Standardization of Travel Chains: The bus data includes multiple fields for each passenger's travel information, such as travel mode, route number, station number, station name, and card swiping time. To simplify the analysis, these fields were integrated into standardized travel chain units by grouping every five fields together, following the passenger ID. This process not only simplified the data structure but also ensured the completeness and clarity of each travel chain. The standardized travel chains are shown in Table~\ref{tab:8}.

% \begin{table}[h]
% \scriptsize % Reduce the font size
% \centering
% \caption{Passengers Travel Chains}
% \label{tab:8}
% \begin{tabularx}{\textwidth}{>{\centering\arraybackslash}X 
%                            >{\centering\arraybackslash}X 
%                            >{\centering\arraybackslash}X
%                            >{\centering\arraybackslash}X}
% \toprule
% \textbf{Passenger ID} & \textbf{Travel Chain1} & \textbf{Travel Chain2} & \textbf{Travel Chain3} \\
% \midrule
% 78aa216c934e339e650f119ced & DT,1,24,SiHui,20180301065200 & DT,Batong Line,11,Liyuan,20180301072757 &  \\
% 78aa216c934e339e650f119ced & DT,1,24,SiHui,20180301082500 & DT,Batong Line,9,Guoyuan,20180301084903 &  \\
% 78aa216c934e339e650f119ced & DT,6,53,Qingnianlu,20180301094300 & DT,6,57,Huangqu,20180301095108 &  \\
% 78aa216c934e339e650f119ced & GJ,675,12,Dougezhuanglu,20180301103001 & GJ,675,17,Qingnianlu Kou,20180301103911 & GJ,2,10,Shilibu,20180301104601 \\
% 78aa216c934e339e650f119ced & DT,6,51,Shilibu,20180301120000 & DT,6,69,Beiyunhe West,20180301123014 &  \\
% 78aa216c934e339e650f119ced & GJ,648,23,Guanzhuang,20180301133201 & GJ,648,24,Zhoujiaojing,20180301135807 &  \\
% \bottomrule
% \end{tabularx}
% \end{table}

\begin{table}[h]
\centering
\caption{Passengers Travel Chains}
\label{tab:8}
% \scriptsize % Reduce font size for better fit
\setlength{\tabcolsep}{5pt} % Adjust column spacing
\renewcommand{\arraystretch}{1.6} % Adjust row spacing
\resizebox{\textwidth}{!}{ % Expand table to full width
\begin{tabular}{p{5cm} p{7cm} p{6cm} p{6cm}} % Adjusted column widths
\toprule
\textbf{Passenger ID} & \textbf{Travel Chain1} & \textbf{Travel Chain2} & \textbf{Travel Chain3} \\
\midrule
78aa216c934e339e650f119ced & DT,1,24,SiHui,20180301065200 & DT,Batong Line,11,Liyuan,20180301072757 &  \\
78aa216c934e339e650f119ced & DT,1,24,SiHui,20180301082500 & DT,Batong Line,9,Guoyuan,20180301084903 &  \\
78aa216c934e339e650f119ced & DT,6,53,Qingnianlu,20180301094300 & DT,6,57,Huangqu,20180301095108 &  \\
78aa216c934e339e650f119ced & GJ,675,12,Dougezhuanglu,20180301103001 & GJ,675,17,Qingnianlu Kou,20180301103911 & GJ,2,10,Shilibu,20180301104601 \\
78aa216c934e339e650f119ced & DT,6,51,Shilibu,20180301120000 & DT,6,69,Beiyunhe West,20180301123014 &  \\
78aa216c934e339e650f119ced & GJ,648,23,Guanzhuang,20180301133201 & GJ,648,24,Zhoujiaojing,20180301135807 &  \\
\bottomrule
\end{tabular}
}
\end{table}

Step 2 - Merging Data: All passenger data within a week was merged and reorganized based on the passenger ID. This consolidation was essential for enabling more efficient data processing in the subsequent analysis.

Step 3 - Determining Travel Frequency: The core of this step involved calculating each passenger's travel frequency within the week. By counting the number of times each passenger ID appears in the dataset per day, the study accurately determined the travel frequency for each passenger over the course of the week.

\begin{figure}[h]
    \centering
    \begin{minipage}{0.48\textwidth}
        \centering
        \includegraphics[width=\linewidth, height=7cm]{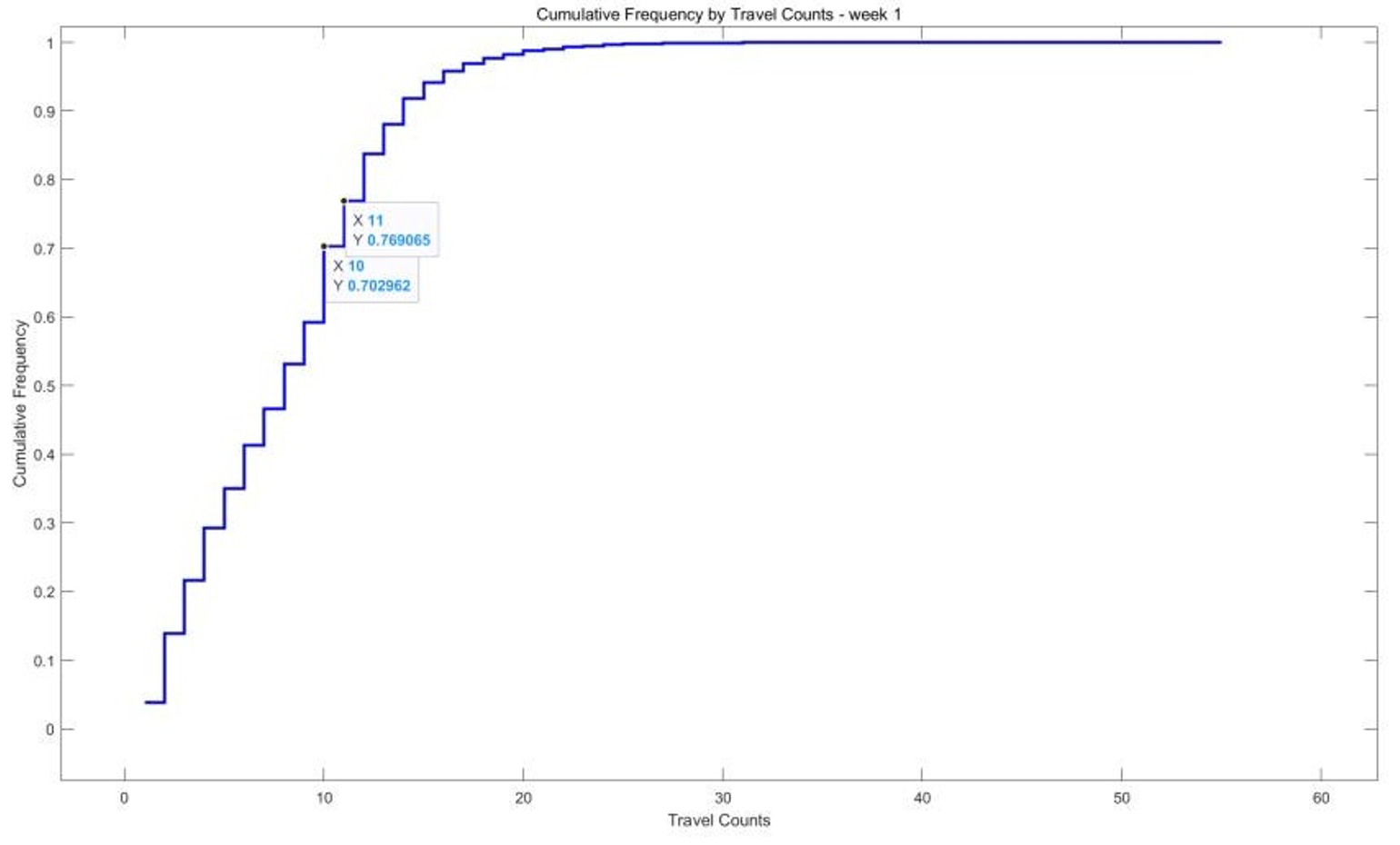}
        \label{fig:41}
    \end{minipage} \hfill
    \begin{minipage}{0.48\textwidth}
        \centering
        \includegraphics[width=\linewidth, height=7cm]{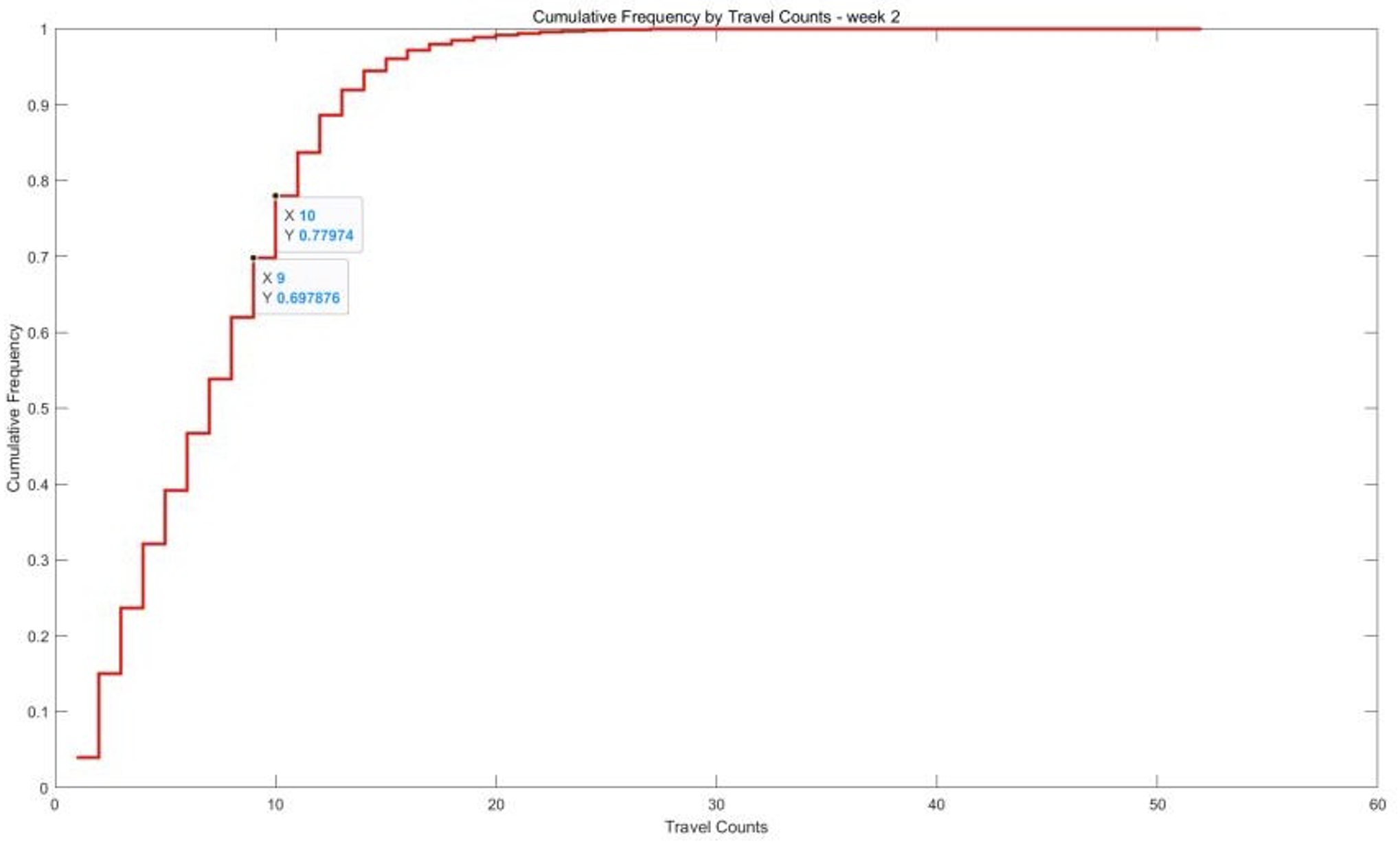}
        \label{fig:42}
    \end{minipage}
    \caption{Cumulative Frequency by travel counts – week 1 (left) and week 2 (right).}
    \label{fig:4}
\end{figure}

Step 4 - Determining the Division Threshold: Through a comprehensive analysis of travel data, the distribution characteristics of travel frequencies were assessed. The study tallied the number of trips and cumulative frequencies over a two-week period, observing the disparity in travel frequencies between high and low-frequency passengers. Figure~\ref{fig:4} illustrates these findings.

Step 5 - Dividing High and Low Frequency Passenger Networks: In the absence of a clear natural boundary in cumulative travel frequencies, a threshold of approximately 25\% was selected. Passengers with the highest 25\% of travel frequencies were classified as high-frequency passengers, while the rest were categorized as low-frequency passengers. Passengers traveling 11 or more times in the first week and 10 or more times in the second week were designated as high-frequency, while all others were classified as low-frequency. The resulting high- and low-frequency passenger networks are presented in Table~\ref{tab:9} and Table~\ref{tab:10}.

\begin{table}[h]
\centering
\caption{Example of High-Frequency Passenger Network for the First Week}
\label{tab:9}
\setlength{\tabcolsep}{5pt} % Adjust column spacing
\renewcommand{\arraystretch}{1.6} % Adjust row spacing
\resizebox{\textwidth}{!}{ % Expand table to full width
\begin{tabular}{p{3cm} p{6cm} p{7cm} p{8cm}} % Adjusted column widths
\toprule
\textbf{Travel Counts} & \textbf{Passenger ID} & \textbf{Travel Chain 1} & \textbf{Travel Chain 2} \\
\midrule
55 & 78aa216c934e339e650f119ced999979 & DT,1,24,SiHui,20180301065200 & DT,Batong Line,11,Liyuan,20180301072757 \\
55 & 78aa216c934e339e650f119ced999979 & DT,1,24,SiHui,20180301082500 & DT,Batong Line,9,Guoyuan,20180301084903 \\
55 & 78aa216c934e339e650f119ced999979 & DT,6,53,Qingnianlu,20180301094300 & DT,6,57,Huangqu,20180301095108 \\
55 & 78aa216c934e339e650f119ced999979 & GJ,675,12,Dougezhuanglu,20180301103001 & GJ,675,17,Qingnianlu Kou,20180301103911 \\
55 & 78aa216c934e339e650f119ced999979 & DT,6,51,Shilibu,20180301120000 & DT,6,69,Beiyunhe West,20180301123014 \\
\bottomrule
\end{tabular}
}
\end{table}

\begin{table}[h]
\centering
\caption{Example of Low-Frequency Passenger Network for the First Week}
\label{tab:10}
\setlength{\tabcolsep}{5pt} % Adjust column spacing
\renewcommand{\arraystretch}{1.6} % Adjust row spacing
\resizebox{\textwidth}{!}{ % Expand table to full width
\begin{tabular}{p{2cm} p{6cm} p{8cm} p{8cm}} % Adjusted column widths
\toprule
\textbf{Travel Counts} & \textbf{Passenger ID} & \textbf{Travel Chain 1} & \textbf{Travel Chain 2} \\
\midrule
10 & 000006c371ae0dd0336028a5149226f & DT,2,0,14,East Gate,20180301185100 & DT,6,0,69,Beiyunhe West,20180301195324 \\
10 & 000006c371ae0dd0336028a5149226f & DT,6,0,75,Dongxiaqu,20180304074800 & DT,2,0,14,East Gate,20180304085100 \\
10 & 000006c371ae0dd0336028a5149226f & DT,2,0,14,East Gate,20180304193100 & DT,6,0,69,Beiyunhe West,20180304020242 \\
10 & 000006c371ae0dd0336028a5149226f & GJ,809,1,3,Small Arts East Area,20180305123401 & GJ,809,1,19,Yunhe Mingzhujiaqu,20180305130025 \\
10 & 000006c371ae0dd0336028a5149226f & DT,2,0,14,East Gate,20180305190100 & DT,6,0,69,Beiyunhe West,20180305195309 \\
\bottomrule
\end{tabular}
}
\end{table}

In the constructed networks, the first week data show that high-frequency passengers had a total of 3,607,960 records, while low-frequency passengers accounted for 9,538,509 records. In the second week, the high-frequency network contained 3,481,020 records, and the low-frequency network had 840,793 records. These comprehensive datasets provide a foundation for analyzing passenger travel behavior and lay the groundwork for further exploration of complex network characteristics.

\subsection{Complex network construction}

In public transport systems, complex networks serve as a powerful tool for studying passenger flows, the connections between stations, and the overall efficiency and robustness of the transport system. In the construction of a complex network for public transport, this study treats stations as nodes, while the travel paths of passengers form the edges connecting these nodes.

The role of the complex network in this study is twofold: It aids in network structure analysis and in examining travel differences between high and low-frequency passengers. By constructing a complex network with stations as nodes, this paper provides an in-depth analysis of the public transportation system's structural characteristics, including the importance of stations (as measured by node degree or intermediary centrality), community structure (based on modularity optimization), and the overall connectivity of the network.

These analyses are critical for identifying key nodes and weak links within the network, offering a scientific foundation for public transport planning and optimization. Furthermore, the construction of complex networks allows for the exploration of travel behavior differences between high and low-frequency passengers at a macroscopic level. By comparing various network types within the high and low-frequency passenger networks, this study aims to reveal differences in spatial distribution and temporal usage patterns between these two groups.

\begin{figure}[h]
    \centering
    \includegraphics[width=10cm, height=9cm]{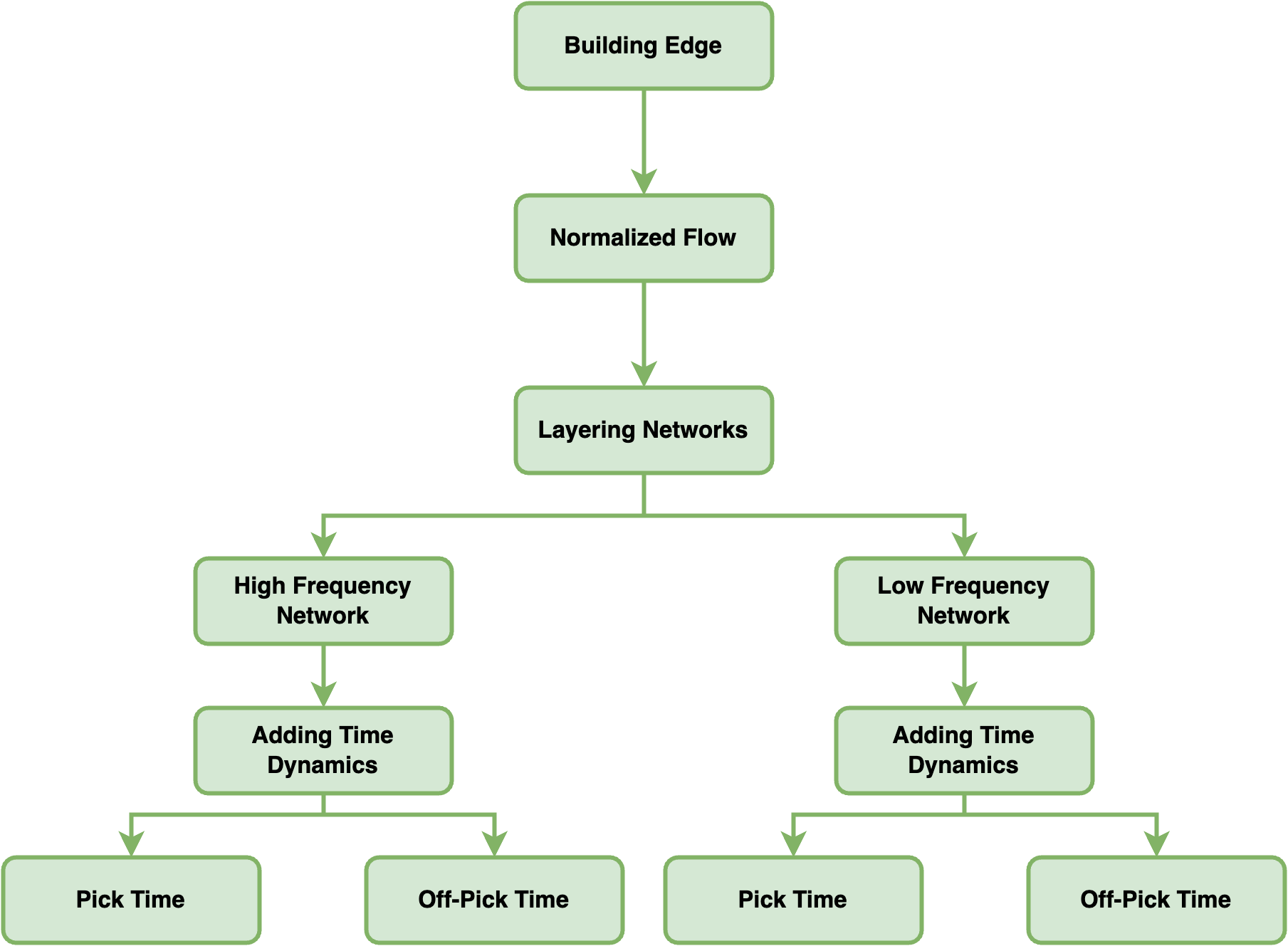}
    \caption{Complex Network Construction Network}
    \label{fig:5}
\end{figure}

In this study, the researchers take a series of steps to construct complex networks, with the primary objective of analyzing the travel differences between high and low-frequency passengers within the public transportation system, as illustrated in Figure~\ref{fig:5}.

Step 1 - Node Identification: All bus and subway stations are treated as nodes within the network. To ensure the accuracy and effectiveness of the network analysis, each station is assigned a unique identifier. In this study, the station names themselves are used directly as identifiers for the nodes.

Step 2 - Edge Construction: The edges connecting the nodes represent passenger flow between stations. An edge is established for every pair of stations where passenger movement is observed, and the flow between these stations is used as the weight of the edge. The weight is calculated based on the frequency of travel between the stations, reflecting the intensity of passenger movement.

Step 3 - Normalization of Flow: Given the varying numbers of passengers in the high and low-frequency networks, max-min normalization is applied to ensure the comparability of station-to-station flow across networks. This technique scales all flow values to a unified range between 0 and 1, preserving the relative position of each value within the maximum and minimum flow values. The specific method for this normalization is presented in Formula 1. 

\begin{equation}
\text{Normalized Flow} = \frac{\text{Flow} - \text{Min Flow}}{\text{Max Flow} - \text{Min Flow}}
\end{equation}

This step eliminates discrepancies caused by differing passenger numbers in the high and low-frequency networks, enabling a fair comparison of flow differences across networks of different scales. It allows for a focus on the network structure itself, such as the distribution patterns of flow, the identification of key connecting stations, and the overall efficiency and robustness of the network. The results of the normalized edges and inter-station flows are shown in Table~\ref{tab:11} and Table~\ref{tab:12}.

\begin{table}[H]
\centering
\small
\caption{Normalized Flow in the High-Frequency Passenger Network for the First Week}
\label{tab:11}
\begin{tabularx}{\textwidth}{>{\centering\arraybackslash}X 
                           >{\centering\arraybackslash}X 
                           >{\centering\arraybackslash}X 
                           >{\centering\arraybackslash}X}
\toprule
\textbf{Original Station} & \textbf{Destination} & \textbf{Flows} & \textbf{Normalized Flow} \\
\midrule
Si Hui Hub Station & Si Hui Hub Station & 16722 & 0.449743 \\
Si Hui & Si Hui Hub Station & 11497 & 0.309207 \\
Da Bei Yao South & Guomao & 7641 & 0.205492 \\
Si Hui Hub Station & Si Hui & 6852 & 0.184271 \\
Ba Wang Fen West & Da Wang Road & 5846 & 0.157212 \\
Si Hui Hub Station & Tongzhou Beiyuan Road East & 5821 & 0.15654 \\
Guomao & Da Bei Yao South & 5600 & 0.150596 \\
Kangjia Gou & Kangjia Gou & 5431 & 0.14605 \\
Da Bei Yao South & Da Bei Yao South & 5157 & 0.13868 \\
Qingnian Road & Qingnian Road North & 4938 & 0.13279 \\
\bottomrule
\end{tabularx}
\end{table}

\begin{table}[H]
\centering
\small
\caption{Normalized Flow in the Low-Frequency Passenger Network for the First Week}
\label{tab:12}
\begin{tabularx}{\textwidth}{>{\centering\arraybackslash}X 
                           >{\centering\arraybackslash}X 
                           >{\centering\arraybackslash}X 
                           >{\centering\arraybackslash}X}
\toprule
\textbf{Original Station} & \textbf{Destination} & \textbf{Flows} & \textbf{Normalized Flow} \\
\midrule
Si Hui Hub Station & Si Hui Hub Station & 37180 & 1 \\
Si Hui & Si Hui Hub Station & 24100 & 0.6481885 \\
Da Bei Yao South & Guomao & 21623 & 0.5815649 \\
Da Bei Yao South & Da Bei Yao South & 20050 & 0.539256 \\
Hongmiao Road East & Hongmiao Road East & 18694 & 0.5027838 \\
Langjia Yuan & Guomao & 16310 & 0.4386616 \\
Da Wang Road & Ba Wang Fen East & 16225 & 0.4363754 \\
Kangjia Gou & Kangjia Gou & 15990 & 0.4300546 \\
Guomao & Langjia Yuan & 15538 & 0.4178972 \\
Guomao & Da Bei Yao South & 14112 & 0.3795422 \\
\bottomrule
\end{tabularx}
\end{table}

Step 4 - Network Hierarchization: The network is subdivided into two sub-networks based on the travel frequency of passengers: high-frequency and low-frequency. This division allows for a more precise analysis and comparison of the travel patterns of different passenger groups, revealing their impact on the structure and fluidity of the public transportation network.

Step 5 - Temporal Dynamics: To account for passenger flow during different time periods, such as peak morning and evening hours or off-peak times, multiple network layers are created to reflect the changing travel demand over time. This temporal dimension allows the network to not only illustrate spatial passenger flow patterns but also capture the effects of temporal variations on these patterns.

\subsection{Visualization of complex networks}
\subsubsection{Spatial Pattern visualization}
To gain a deeper understanding of the travel preferences of high- and low-frequency passengers and their impact on the transportation network, the top 1,000 OD (origin-destination) pairs for each passenger group were ranked and compared. The rankings for these OD pairs were calculated separately for high-frequency and low-frequency passenger networks during peak and off-peak periods. Subsequently, the ranking differences were determined by subtracting the low-frequency ranking from the high-frequency ranking. For example, if a route ranks first in the high-frequency network but 1,000th in the low-frequency network, the difference is -999. If an OD pair is absent in the low-frequency network, its ranking difference is assigned a value of -1,000.

This approach provides a detailed perspective on how certain routes are predominantly used by specific user groups and are less significant for others. These differences can be attributed to passenger travel purposes or habits. For instance, high-frequency passengers might favor efficient or convenient routes for commuting, while low-frequency passengers may rely on different routes for occasional or leisure travel. 

\begin{figure}[h]
    \centering
    \includegraphics[width=10cm, height=8cm]{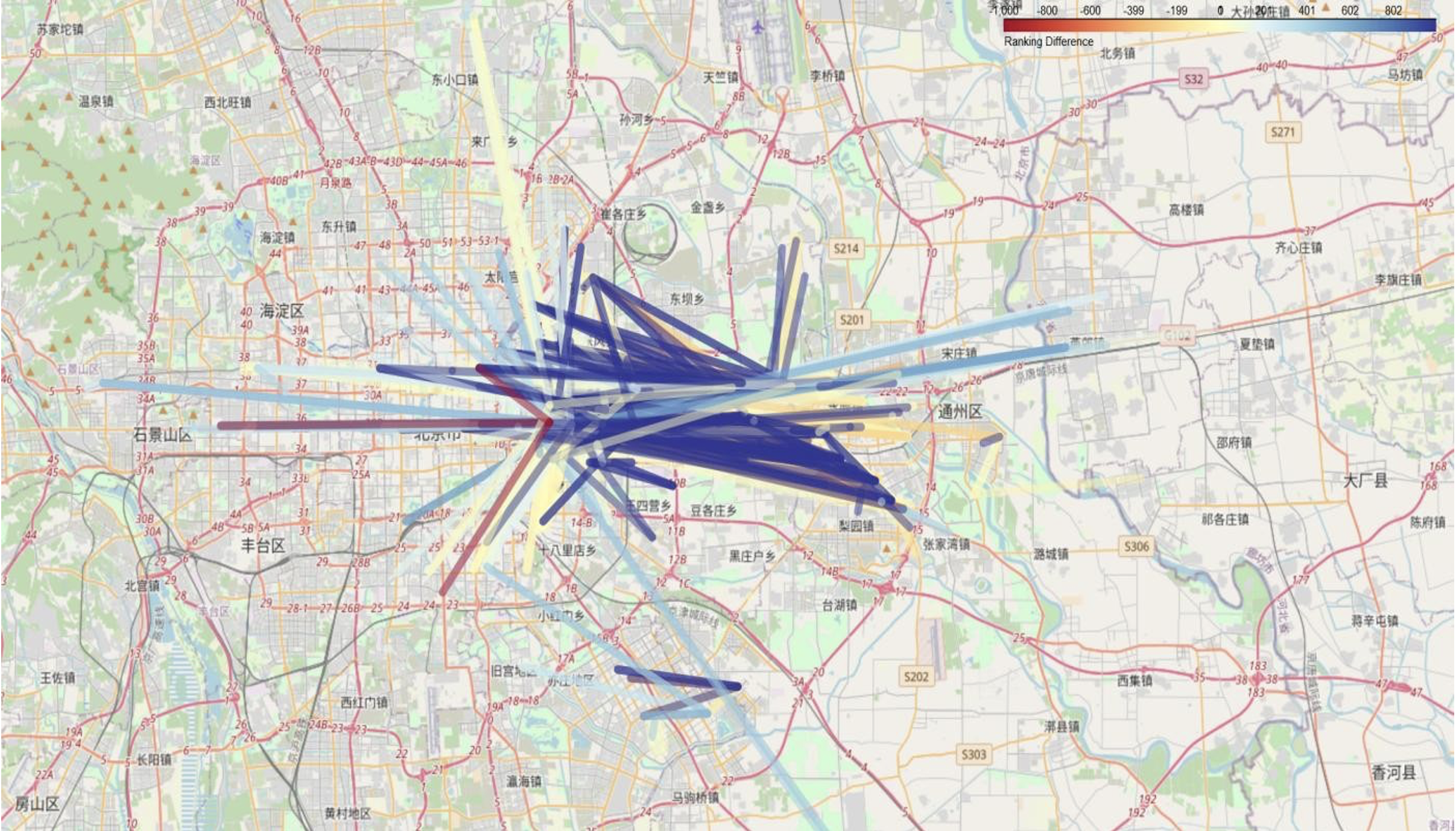}
    \caption{High and low frequency passenger flow chart for the first week}
    \label{fig:6}
\end{figure}

\begin{figure}[h]
    \centering
    \includegraphics[width=10cm, height=8cm]{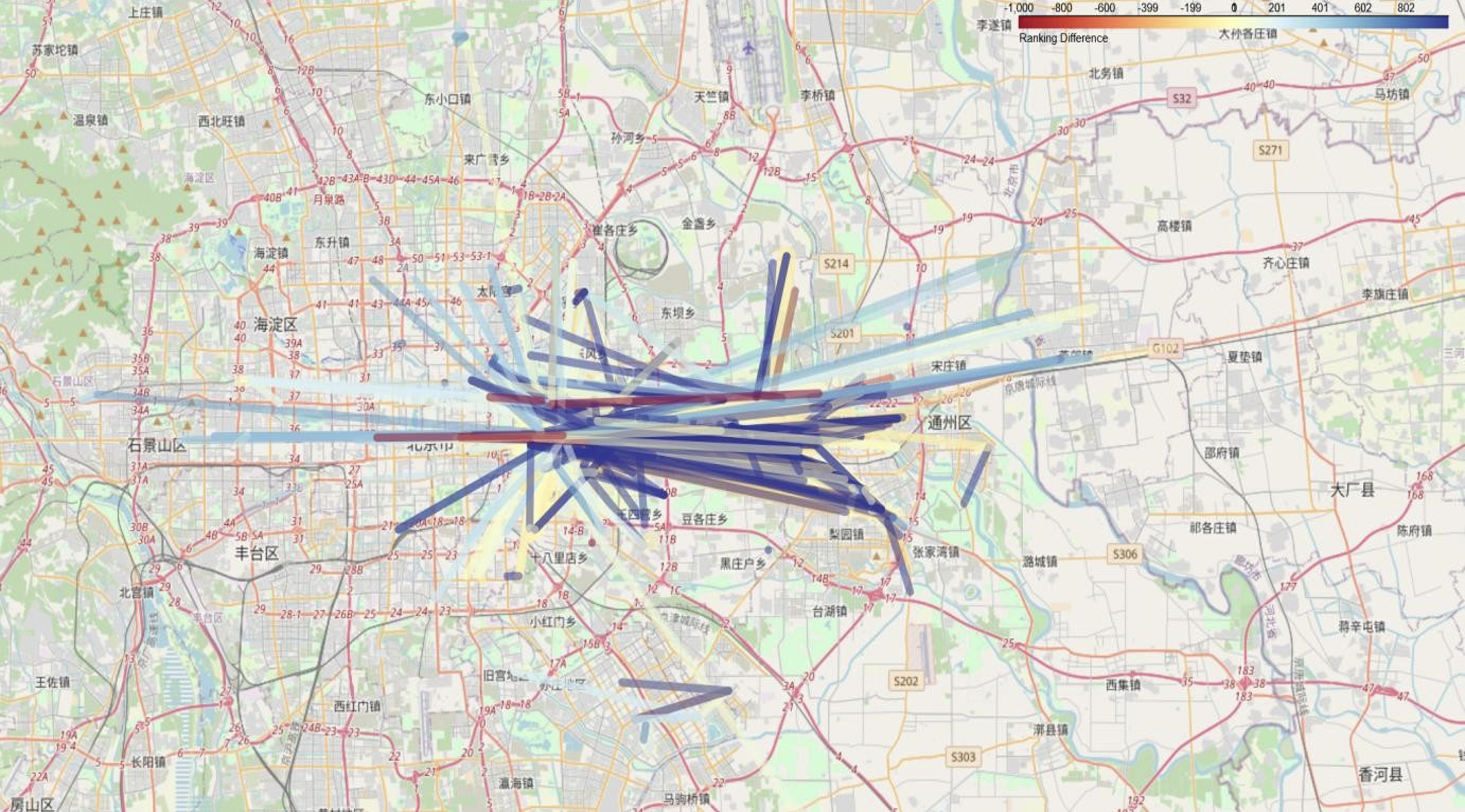}
    \caption{High and low frequency passenger flow chart for the second week}
    \label{fig:7}
\end{figure}

To visualize Table~\ref{tab:13}, the flow of OD pairs is displayed in different colors according to the ranking difference, corresponding to the upper right legend of Figure~\ref{fig:6} and Figure~\ref{fig:7}.

\begin{table}[H]
\centering
\scriptsize
\caption{OD Pairs of High and Low Frequency Networks in the First Week}
\label{tab:13}
\begin{tabularx}{\textwidth}{>{\centering\arraybackslash}X 
                           >{\centering\arraybackslash}X 
                           >{\centering\arraybackslash}X}
\toprule
\textbf{Start-station} & \textbf{Terminate-station} & \textbf{Rank Difference} \\
\midrule
Baliqiao & Tongzhou Yangzhuang North & 1000 \\
Jishuitan & Deshengmen West & 811 \\
Dongzhimen & Zuojiazhuang & 704 \\
Beijing South Station & Da Wang Road & 664 \\
Xingda Square Community & Ba Wang Fen West & 640 \\
Dongzhimen Hub Station & Dongzhimen & 628 \\
Shuangjing Bridge North & Da Bei Yao South & 623 \\
Zhongzhaofu Village & Langjia Yuan & 614 \\
Langjia Yuan & Xingda Square Community & 614 \\
Guomao & Wukesong & 613 \\
\bottomrule
\end{tabularx}
\end{table}

\subsubsection{Temporal Flow Visualization}

In this study, as illustrated in Table~\ref{tab:14}, the travel records of each passenger are marked by identifying their initial card-swiping event as the starting point to construct their travel chain dataset for specific time slots. For example, the passenger with the ID "78aa216c934e339e650f119ced699979" performed their first card swipe at several distinct times on March 1, 2018, each corresponding to a different travel chain. The initial card-swiping event serves as a precise timestamp for each travel chain, enabling the systematic compilation of all passengers' first card-swiping events within each defined time slot.

This compilation forms a robust dataset that captures the flow of passengers during specific time periods. Such data provides the foundation for further analysis, enabling a detailed examination of travel patterns and identifying peak period flows. This approach is essential for understanding temporal dynamics in the public transportation system and offers valuable insights for optimizing service schedules and resource allocation.

\begin{table}[H]
\centering
\small
\caption{Example of Passenger Travel Chains}
\label{tab:14}
\begin{tabularx}{\textwidth}{>{\centering\arraybackslash}X 
                           >{\centering\arraybackslash}X 
                           >{\centering\arraybackslash}X}
\toprule
\textbf{Travel Counts} & \textbf{Passenger ID} & \textbf{Travel Chain 1} \\
\midrule
55 & 78aa216c934e339e650f & DT,1,24,Si Hui,20180301065200 \\
55 & 78aa216c934e339e650f & DT,1,24,Si Hui,20180301082500 \\
55 & 78aa216c934e339e650f & DT,6,53,Qingnian Road,20180301094300 \\
55 & 78aa216c934e339e650f & GJ,675,12,Dougezhuang South,20180301103001 \\
55 & 78aa216c934e339e650f & DT,6,51,Shilibao,20180301120000 \\
55 & 78aa216c934e339e650f & DT,1,24,Si Hui,20180301065200 \\
55 & 78aa216c934e339e650f & DT,1,24,Si Hui,20180301082500 \\
\bottomrule
\end{tabularx}
\end{table}

The described method allows for recording the total number of card swipes within each time slot, providing a detailed view of passenger activity during specific periods. By further aggregating the number of swipes across all days of the week for each time slot, the total passenger flow for each time slot throughout the entire week can be determined. This aggregation provides a comprehensive dataset for analyzing weekly travel patterns and identifying peak usage periods. The summarized results are presented in Table Table~\ref{tab:15}.

\begin{table}[H]
\centering
\small
\caption{Total Passenger Flow for Each Time Slot Throughout the Week}
\label{tab:15}
\begin{tabularx}{\textwidth}{>{\centering\arraybackslash}X 
                           >{\centering\arraybackslash}X 
                           >{\centering\arraybackslash}X 
                           >{\centering\arraybackslash}X 
                           >{\centering\arraybackslash}X}
\toprule
\textbf{Time Slot} & \textbf{High-Frequency Week 1} & \textbf{Low-Frequency Week 1} & \textbf{High-Frequency Week 2} & \textbf{Low-Frequency Week 2} \\
\midrule
00-02 & 873 & 1689 & 923 & 1508 \\
02-04 & 481 & 994 & 541 & 797 \\
04-06 & 31984 & 75706 & 45361 & 71866 \\
06-08 & 574449 & 1128045 & 577760 & 1021976 \\
08-10 & 685076 & 1538264 & 642371 & 1428009 \\
10-12 & 255185 & 792695 & 251382 & 784971 \\
12-14 & 260506 & 819192 & 262857 & 839557 \\
14-16 & 265715 & 878660 & 267877 & 870069 \\
16-18 & 492632 & 1328076 & 485278 & 1213996 \\
18-20 & 661251 & 1313580 & 605854 & 1165313 \\
20-22 & 305452 & 544129 & 273537 & 529395 \\
22-24 & 74355 & 117478 & 67278 & 113334 \\
\bottomrule
\end{tabularx}
\end{table}

As illustrated in Figure 8, the passenger flow distribution reveals two distinct peaks in the daily travel cycle: the morning peak (06:00–08:00) and the evening peak (18:00–20:00). During these periods, passenger flow increases significantly, highlighting the concentrated travel demand. This periodic fluctuation provides a clear basis for defining the morning and evening peak periods.

\begin{figure}[h]
    \centering
    \includegraphics[width=10cm, height=8cm]{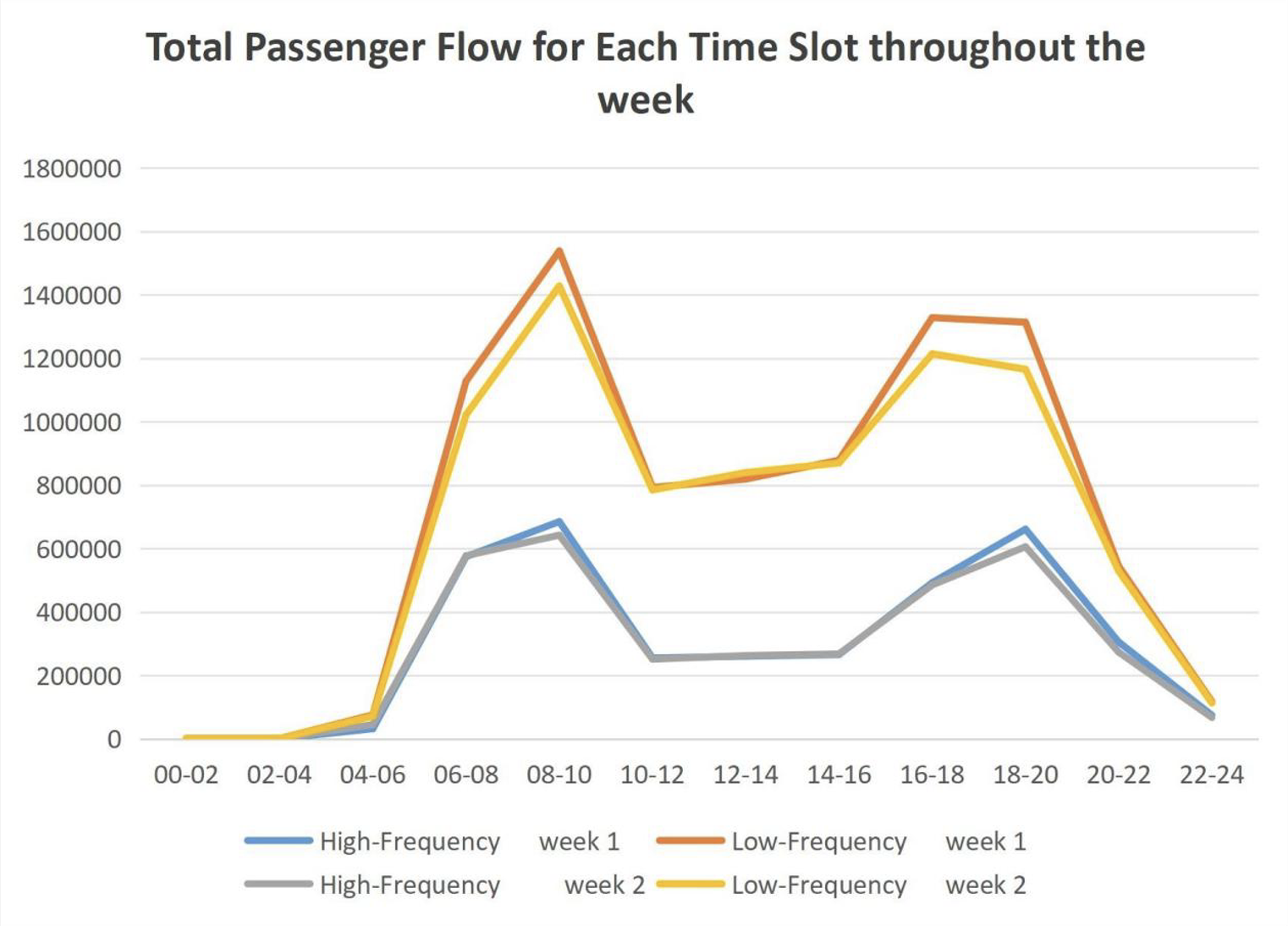}
    \caption{Total Passenger Flow for Each Time Slot throughout the week}
    \label{fig:8}
\end{figure}

Given the differences in the size of high- and low-frequency passenger groups, it is essential to standardize the passenger flow data to ensure accurate comparisons. The standardized data, presented in Table~\ref{tab:16}, reflects the distribution characteristics of passenger flow in the two networks more effectively, enabling a detailed examination of travel behaviors and patterns between high- and low-frequency passenger groups. This standardized approach allows for a clearer understanding of the unique dynamics within each network.

\begin{table}[H]
\centering
\small
\caption{Normalized Passenger Flow for Each Time Slot Throughout the Week}
\label{tab:16}
\begin{tabularx}{\textwidth}{>{\centering\arraybackslash}X 
                           >{\centering\arraybackslash}X 
                           >{\centering\arraybackslash}X 
                           >{\centering\arraybackslash}X 
                           >{\centering\arraybackslash}X}
\toprule
\textbf{Time Interval} & \textbf{Normalized High-Frequency Week 1} & \textbf{Normalized Low-Frequency Week 1} & \textbf{Normalized High-Frequency Week 2} & \textbf{Normalized Low-Frequency Week 2} \\
\midrule
00-02 & 0.000254912 & 0.000785546 & 0.000287427 & 0.000667845 \\
02-04 & 0 & 0.000333597 & 0.0000390172 & 0.000205491 \\
04-06 & 0.020485985 & 0.048917825 & 0.029184872 & 0.046420724 \\
06-08 & 0.373243819 & 0.733239996 & 0.375396919 & 0.664264724 \\
08-10 & 0.445183098 & 1 & 0.4174126 & 0.928302628 \\
10-12 & 0.165630651 & 0.515166314 & 0.163157611 & 0.510143499 \\
12-14 & 0.169090828 & 0.532396964 & 0.170619652 & 0.545640055 \\
14-16 & 0.172478171 & 0.57106822 & 0.173884092 & 0.565481606 \\
16-18 & 0.320039303 & 0.863317516 & 0.315257094 & 0.789132797 \\
18-20 & 0.429690015 & 0.853890959 & 0.393666076 & 0.757474884 \\
20-22 & 0.198318618 & 0.353527123 & 0.177564715 & 0.343945797 \\
22-24 & 0.048039288 & 0.076081606 & 0.043437208 & 0.073386817 \\
\bottomrule
\end{tabularx}
\end{table}

\begin{figure}[h]
    \centering
    \includegraphics[width=10cm, height=8cm]{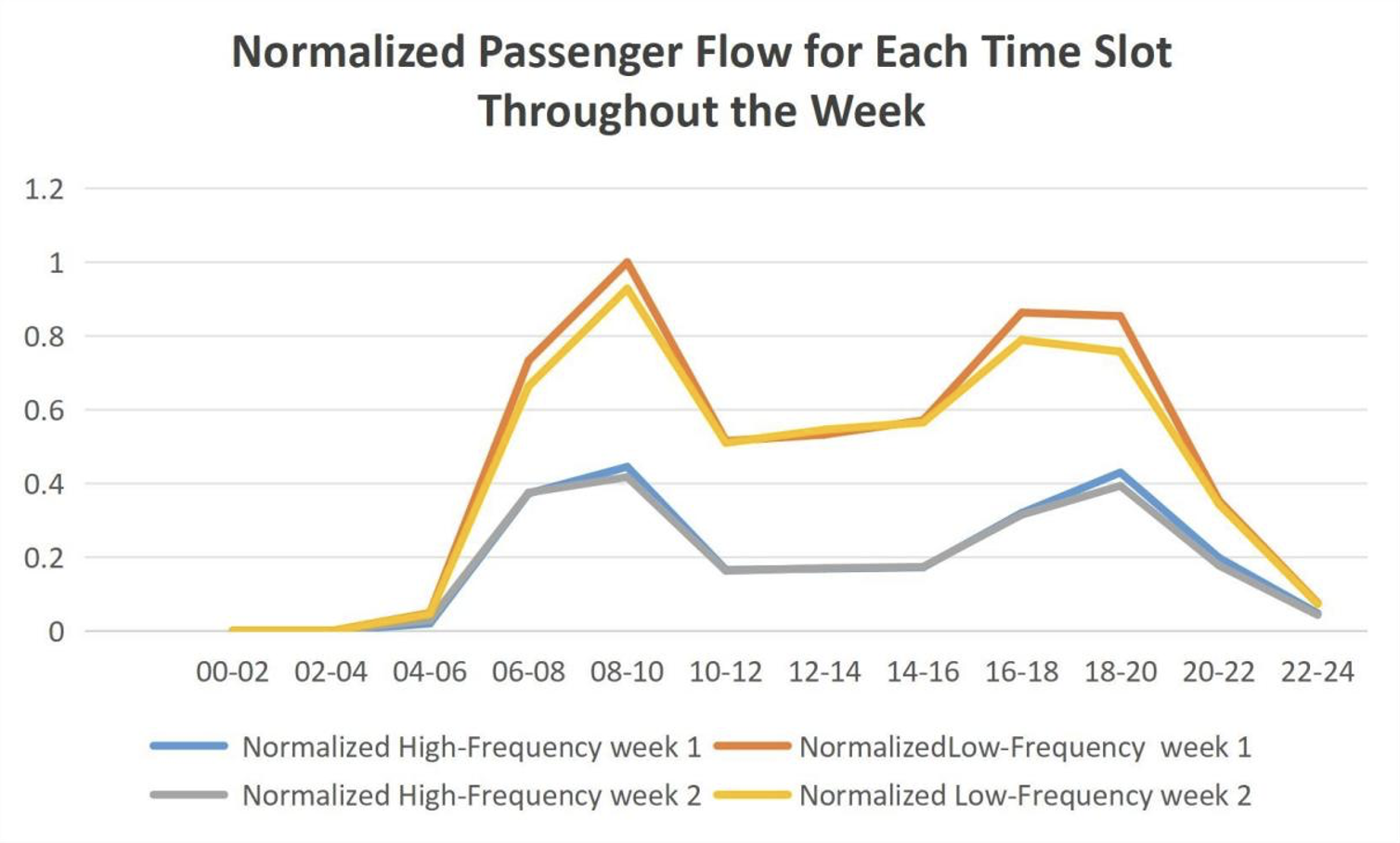}
    \caption{Standardized Passenger Flow for Each Time}
    \label{fig:9}
\end{figure}

Analyzing Figure~\ref{fig:9} yields several important insights into the travel behaviors of high- and low-frequency passengers:

(1) Figure 10 highlights that the orange and yellow curves, representing low-frequency passengers, consistently show a higher flow ratio compared to the high-frequency passenger curves, indicated by the blue and gray lines. This pattern correlates with the overall distribution of passenger numbers, where low-frequency passengers outnumber high-frequency passengers by approximately three to one. However, the flow ratio for low-frequency passengers is only about twice as much, suggesting a relatively higher usage intensity by high-frequency passengers.

(2) Across different times of the day, the travel trends of high- and low-frequency passengers exhibit similar patterns, particularly during the morning peak (8:00–10:00) and evening peak (18:00–20:00), where both groups reach maximum flow levels. However, a significant divergence is observed during the evening peak period, where the flow of low-frequency passengers stabilizes or slightly declines, while high-frequency passenger flow increases. This distinction may reflect the specific activities or work routines of high-frequency passengers, such as extended work hours or evening commutes.

(3) Over the course of the day, the flow difference between high- and low-frequency passengers gradually widens, peaking during the morning and evening rush hours (8:00–10:00). This widening gap could indicate a higher dependency on public transportation among high-frequency passengers during these critical periods. However, during nighttime hours, the flow difference between the two groups becomes almost negligible, suggesting that under conditions of lower overall travel demand, travel frequency plays a less significant role in shaping flow patterns. This observation underscores the importance of optimizing nighttime public transportation services, with a focus on addressing the specific travel needs of both passenger groups.

These findings provide a nuanced understanding of passenger behavior and can guide transportation planners in optimizing services for different passenger segments, particularly during peak and off-peak periods.

\subsubsection{Integrated Spatiotemporal Visualization}

Following the time analysis in the previous section, it is evident that the travel peaks occur between 8:00–10:00 AM and 18:00–20:00 PM. Based on these peak times, this section provides a spatio-temporal visual analysis of the station flow during these periods. The station flow data for the morning and evening peaks are shown in Table~\ref{tab:17} and Table~\ref{tab:18}.

\begin{table}[H]
\centering
\scriptsize
\caption{High-Frequency Passenger Flow Between the Morning Rush Hour Stations in the First Week}
\label{tab:17}
\begin{tabularx}{\textwidth}{>{\centering\arraybackslash}X 
                           >{\centering\arraybackslash}X 
                           >{\centering\arraybackslash}X}
\toprule
\textbf{Start-station} & \textbf{Terminate-station} & \textbf{Flow} \\
\midrule
Da Bei Yao South & Guomao & 4962 \\
Ba Wang Fen West & Da Wang Road & 3581 \\
Si Hui Hub Station & Si Hui & 3125 \\
Tuqiao Village & Tuqiao & 3066 \\
Si Hui East Station & Si Hui East & 2493 \\
\bottomrule
\end{tabularx}
\end{table}

\begin{table}[H]
\centering
\scriptsize
\caption{High-Frequency Passenger Flow Between the Evening Rush Hour Stations in the First Week}
\label{tab:18}
\begin{tabularx}{\textwidth}{>{\centering\arraybackslash}X 
                           >{\centering\arraybackslash}X 
                           >{\centering\arraybackslash}X}
\toprule
\textbf{Start-station} & \textbf{Terminate-station} & \textbf{Flow} \\
\midrule
Si Hui & Si Hui Hub Station & 4456 \\
Guomao & Da Bei Yao South & 2918 \\
Si Hui Hub Station & Tongzhou Beiyuan Road East & 2729 \\
Guan Zhuang & Yangzha Road South & 2417 \\
Guomao & Langjia Yuan & 1983 \\
\bottomrule
\end{tabularx}
\end{table}

\begin{figure}[h]
    \centering
    \begin{minipage}{0.48\textwidth}
        \centering
        \includegraphics[width=\linewidth, height=7cm]{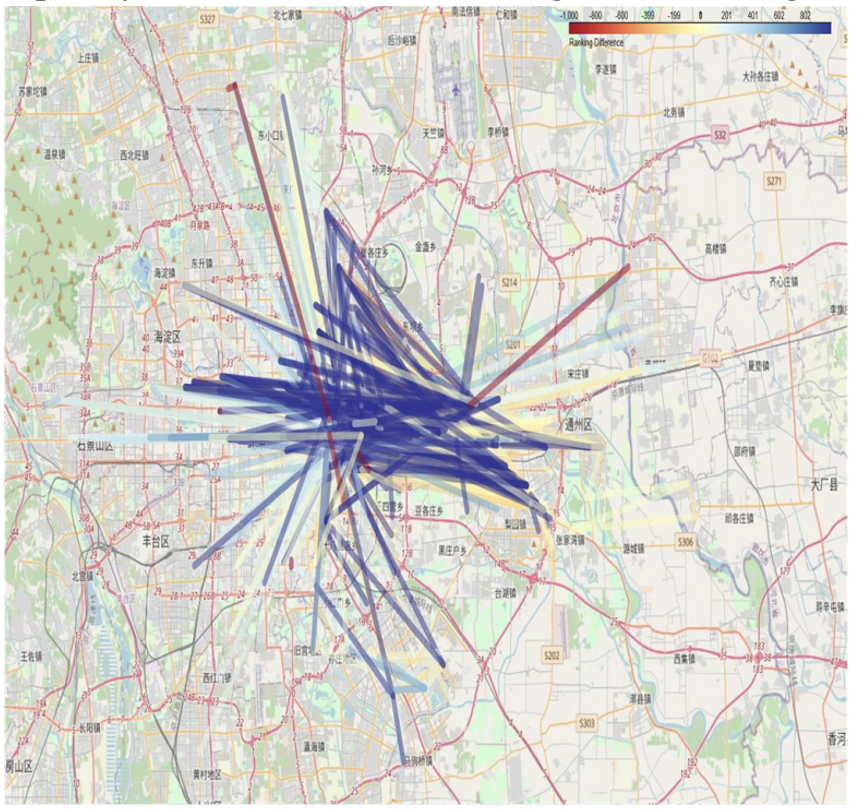}
        \label{fig:101}
    \end{minipage} \hfill
    \begin{minipage}{0.48\textwidth}
        \centering
        \includegraphics[width=\linewidth, height=7cm]{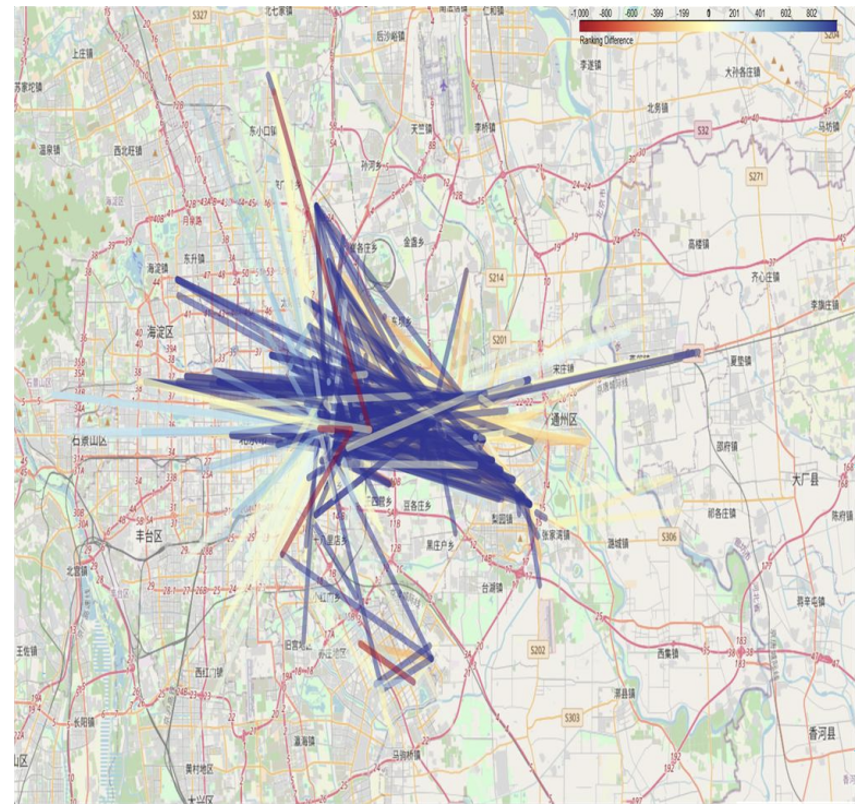}
        \label{fig:102}
    \end{minipage}
    \caption{Comparison of morning peak (left) and evening peak (right) for High frequency network in week 1.}
    \label{fig:10}
\end{figure}

\begin{figure}[h]
    \centering
    \begin{minipage}{0.48\textwidth}
        \centering
        \includegraphics[width=\linewidth, height=7cm]{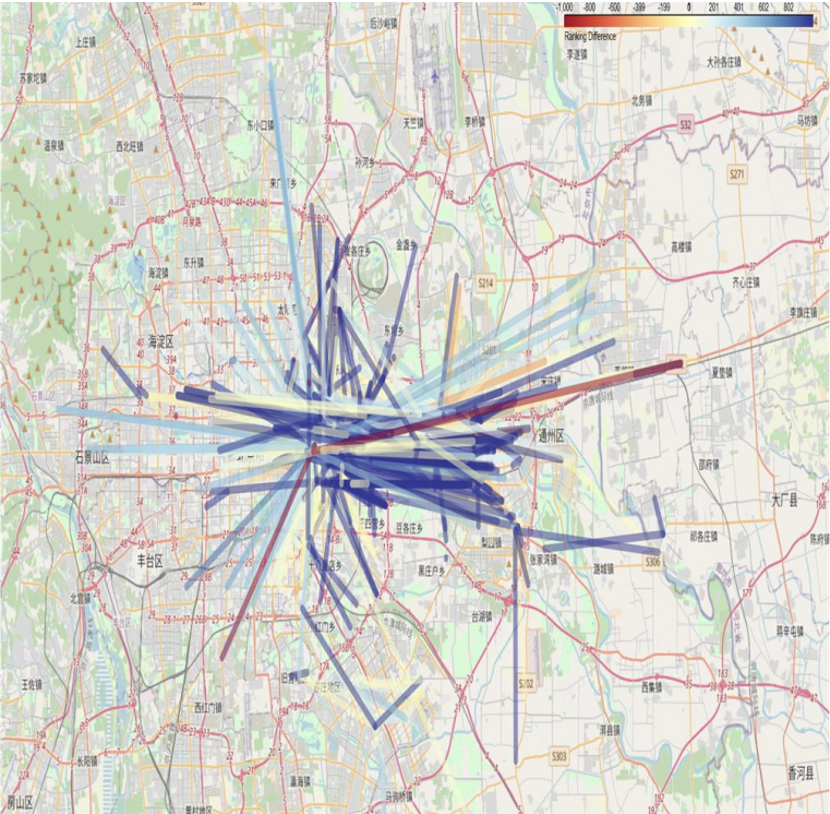}
        \label{fig:111}
    \end{minipage} \hfill
    \begin{minipage}{0.48\textwidth}
        \centering
        \includegraphics[width=\linewidth, height=7cm]{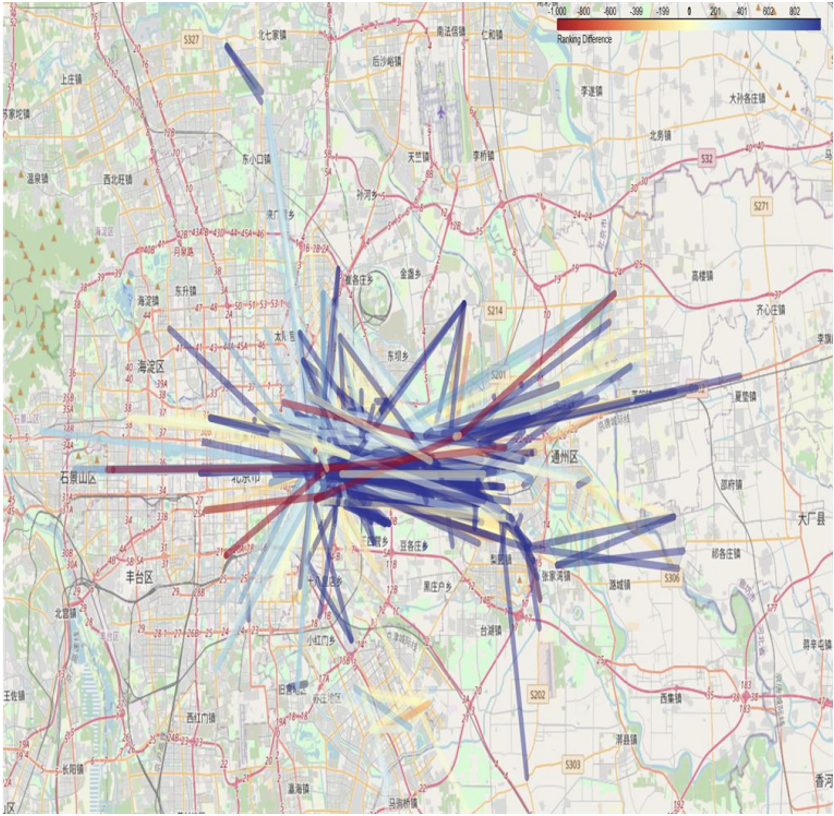}
        \label{fig:112}
    \end{minipage}
    \caption{Comparison of morning peak (left) and evening peak (right) for Low frequency network in week 1.}
    \label{fig:11}
\end{figure}

\begin{figure}[h]
    \centering
    \begin{minipage}{0.48\textwidth}
        \centering
        \includegraphics[width=\linewidth, height=7cm]{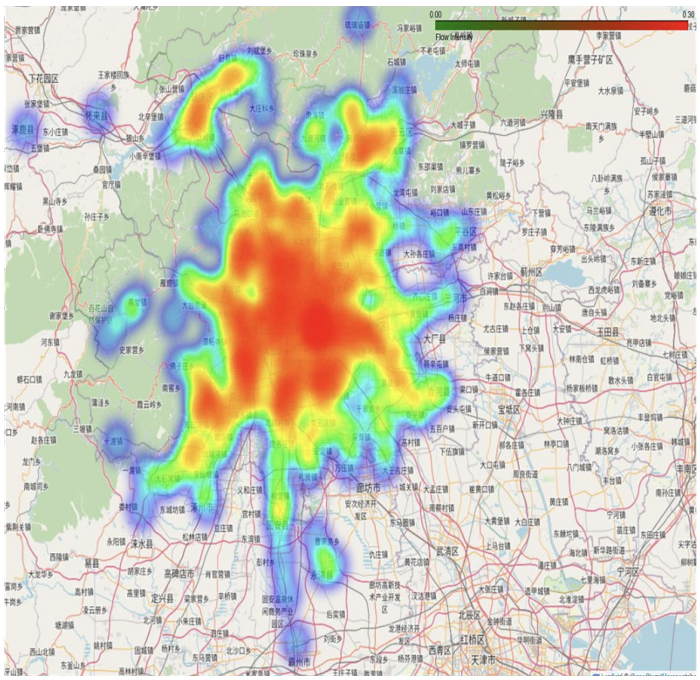}
        \label{fig:121}
    \end{minipage} \hfill
    \begin{minipage}{0.48\textwidth}
        \centering
        \includegraphics[width=\linewidth, height=7cm]{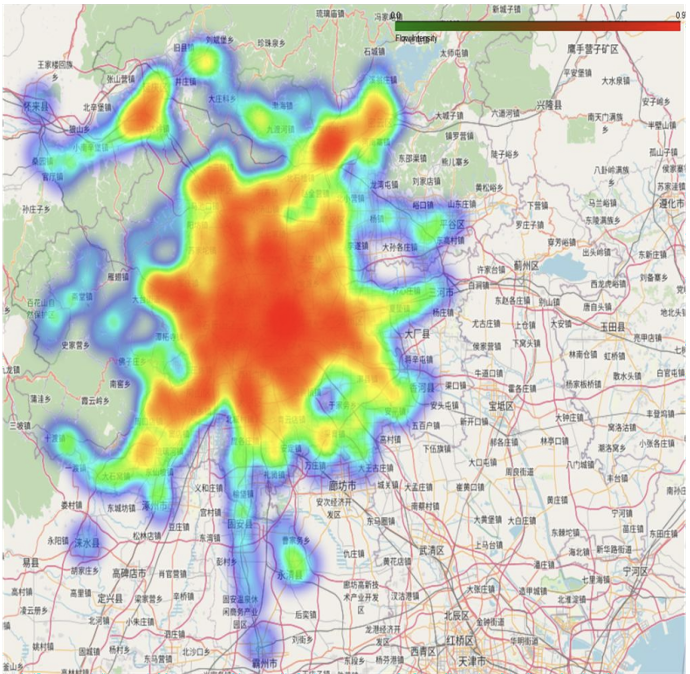}
        \label{fig:122}
    \end{minipage} 
    \caption{High frequency (left) and Low Frequency (right) passenger network in week 1.}
    \label{fig:12}
\end{figure}

\begin{figure}[h]
    \centering
    \begin{minipage}{0.48\textwidth}
        \centering
        \includegraphics[width=\linewidth, height=7cm]{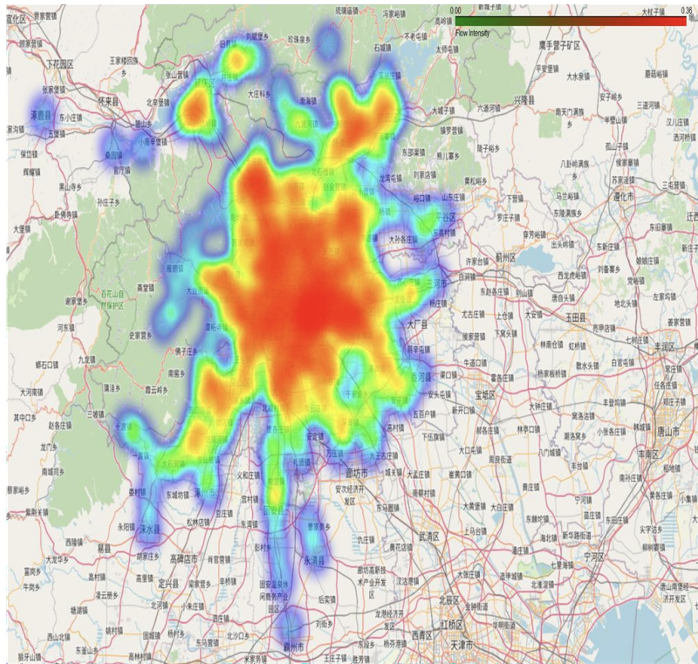}
        \label{fig:131}
    \end{minipage} \hfill
    \begin{minipage}{0.48\textwidth}
        \centering
        \includegraphics[width=\linewidth, height=7cm]{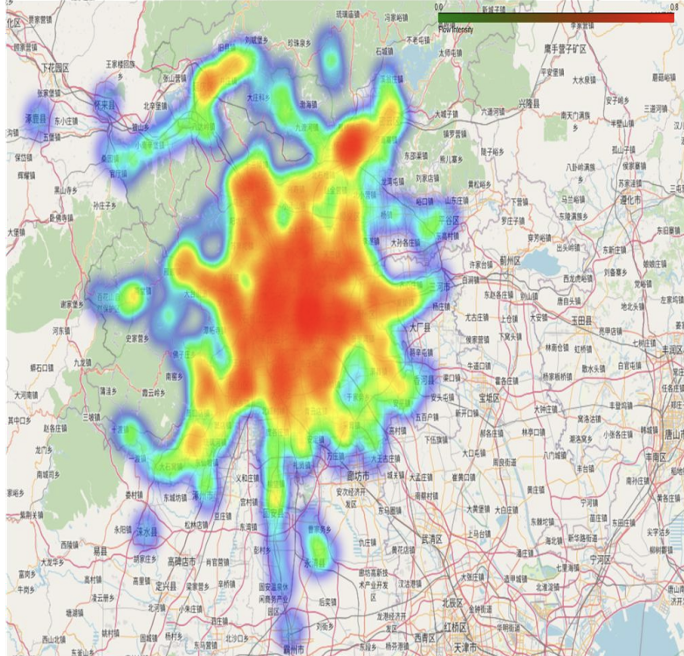}
        \label{fig:132}
    \end{minipage}
    \caption{High frequency (left) and Low Frequency (right) passenger network in week 2.}
    \label{fig:13}
\end{figure}

Approximately 130,000 data points from the morning and evening peaks of the high-frequency passenger network and around 190,000 inter-station traffic data points from the low-frequency network during the same periods were collected. Using the same visualization approach outlined in Section 3.3.1, maps depicting the morning and evening peak hours for both high- and low-frequency passenger networks were created. The visualization focuses on the 1,000 busiest routes, with line color contrasts and thickness emphasizing the frequency of travel. These results are illustrated in Figures~\ref{fig:10}, \ref{fig:11}, \ref{fig:12}, and \ref{fig:13}.

\subsubsection{Cluster Visualization }

This study constructs a daily travel chain network for bus passengers, representing stations as nodes and passenger travel chains as edges. Using a unified storage format, the structured database allows easy retrieval of travel trajectories and detailed analysis of the spatial and temporal characteristics of bus trips.

High- and low-frequency passenger networks are also developed by distinguishing station usage frequencies. The high-frequency network highlights frequently used connections, while the low-frequency network captures less utilized links. This approach enables a detailed exploration of travel behavior and supports the precise optimization of the bus network.

Visualization extends beyond graphical representation, incorporating analysis to reveal differences between high- and low-frequency networks. By comparing standardized flow maps across time periods, peaks, valleys, and distinct activity patterns of passenger groups are observed, offering insights for better network planning.

\section{Complex Network Characteristics Analysis}

\subsection{Node analysis}
A comprehensive analysis of node degree, Betweenness centrality, and closeness centrality provides a deep understanding of the structural characteristics and operational status of the transportation network. This approach helps identify key nodes, referred to as central stations, which are marked by high node degree, high Betweenness centrality, and high closeness centrality. These central nodes hold strategic significance within the network, serving as critical hubs for connectivity and flow management.

The node degree correlation indices of the network are presented in Table~\ref{tab:19}. In each box plot, the bottom and top of the box represent the first quartile (Q1) and the third quartile (Q3), respectively, while the line within the box indicates the median. Points outside the box extend to the minimum and maximum of the data distribution or are located beyond 1.5 times the interquartile range, which are considered potential outliers and may represent special cases or anomalies.

An average line, depicted in red, is included to provide an additional measure of central tendency. To evaluate whether statistically significant differences exist among the same index groups across different network states, the Mann-Whitney U test is conducted. The p-value results from this test are used to determine whether the "no difference" hypothesis can be rejected, with a p-value less than 0.05 indicating significant differences.

\begin{table}[H]
\centering
\small
\caption{Example of the Centrality Index of the High-Frequency Network Nodes in the First Week}
\label{tab:19}
\begin{tabularx}{\textwidth}{>{\centering\arraybackslash}X 
                           >{\centering\arraybackslash}X 
                           >{\centering\arraybackslash}X 
                           >{\centering\arraybackslash}X}
\toprule
\textbf{Node} & \textbf{Node Degree} & \textbf{Betweenness Centrality} & \textbf{closeness centrality} \\
\midrule
234 & 584 & 0.032328503 & 0.72360253 \\
121 & 576 & 0.02718349 & 0.720026241 \\
431 & 519 & 0.017041975 & 0.686116057 \\
245 & 516 & 0.021768721 & 0.677606091 \\
470 & 505 & 0.012482327 & 0.672393736 \\
42  & 492 & 0.013863119 & 0.662205952 \\
292 & 487 & 0.022916277 & 0.670331179 \\
392 & 486 & 0.021179668 & 0.669304638 \\
154 & 478 & 0.012861307 & 0.647490264 \\
357 & 467 & 0.011550909 & 0.656240133 \\
\bottomrule
\end{tabularx}
\end{table}

When comparing the node degree distribution of the high-frequency passenger complex network over two time periods, the results reveal significant statistical differences. The P-value for group 01 is 0.022, and the P-value for group 02 is 0.033, both of which are below the traditional significance threshold of 0.05. This indicates that there are statistically significant differences in the node degree distribution between the high-frequency and low-frequency networks across the two periods.

\begin{figure}[h]
    \centering
    \includegraphics[width=9cm, height=7cm]{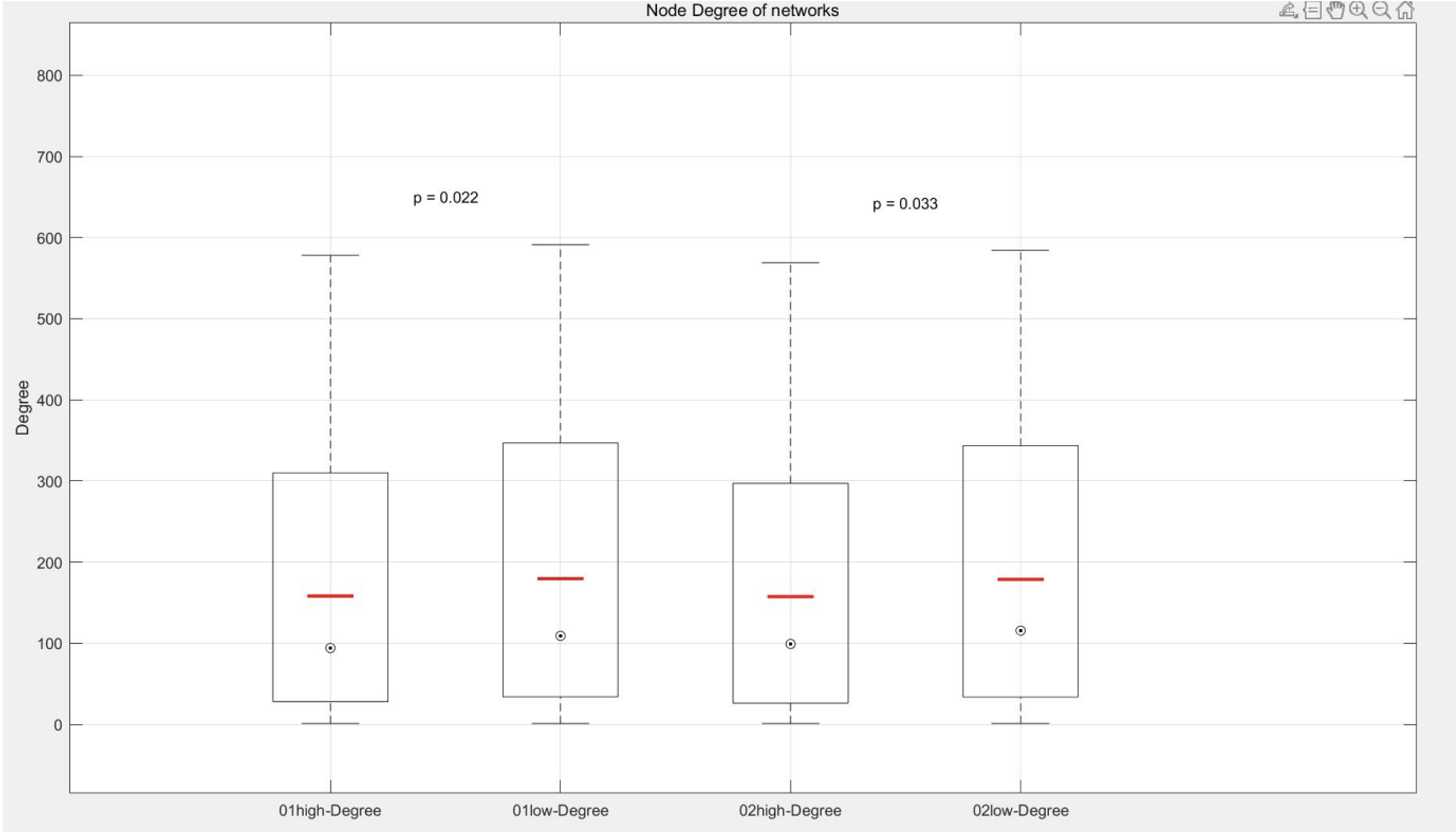}
    \caption{Box plots of node degree}
    \label{fig:14}
\end{figure}

As shown in Figure~\ref{fig:14}, the median node degree is generally lower in the high-frequency network than in the low-frequency network. This finding suggests that the sites in low-frequency networks are typically connected to a larger number of other sites, forming a more active network of traffic nodes. This characteristic reflects the central role of certain stations in the low-frequency network, such as major transportation hubs or commercial centers, which exhibit higher node degrees due to serving a larger volume of passengers or connecting more lines.

In contrast, the lower median node degree in high-frequency networks may indicate that the stations primarily used by high-frequency passengers tend to occupy more peripheral or marginal positions in the overall station network. This distinction underscores the differing roles and spatial characteristics of stations in high- and low-frequency passenger networks.

When comparing the betweenness centrality distribution of high-frequency and low-frequency passenger networks across the two periods, the observed p-values are 0.708 and 0.929, respectively. Since these values are significantly above the standard significance threshold of 0.05, it can be concluded that there is no statistically significant difference in Betweenness centrality between the two networks. This indicates that the importance of stations as passenger traffic transfer points remains relatively consistent in both high-frequency and low-frequency networks.

\begin{figure}[h]
    \centering
    \includegraphics[width=9cm, height=7cm]{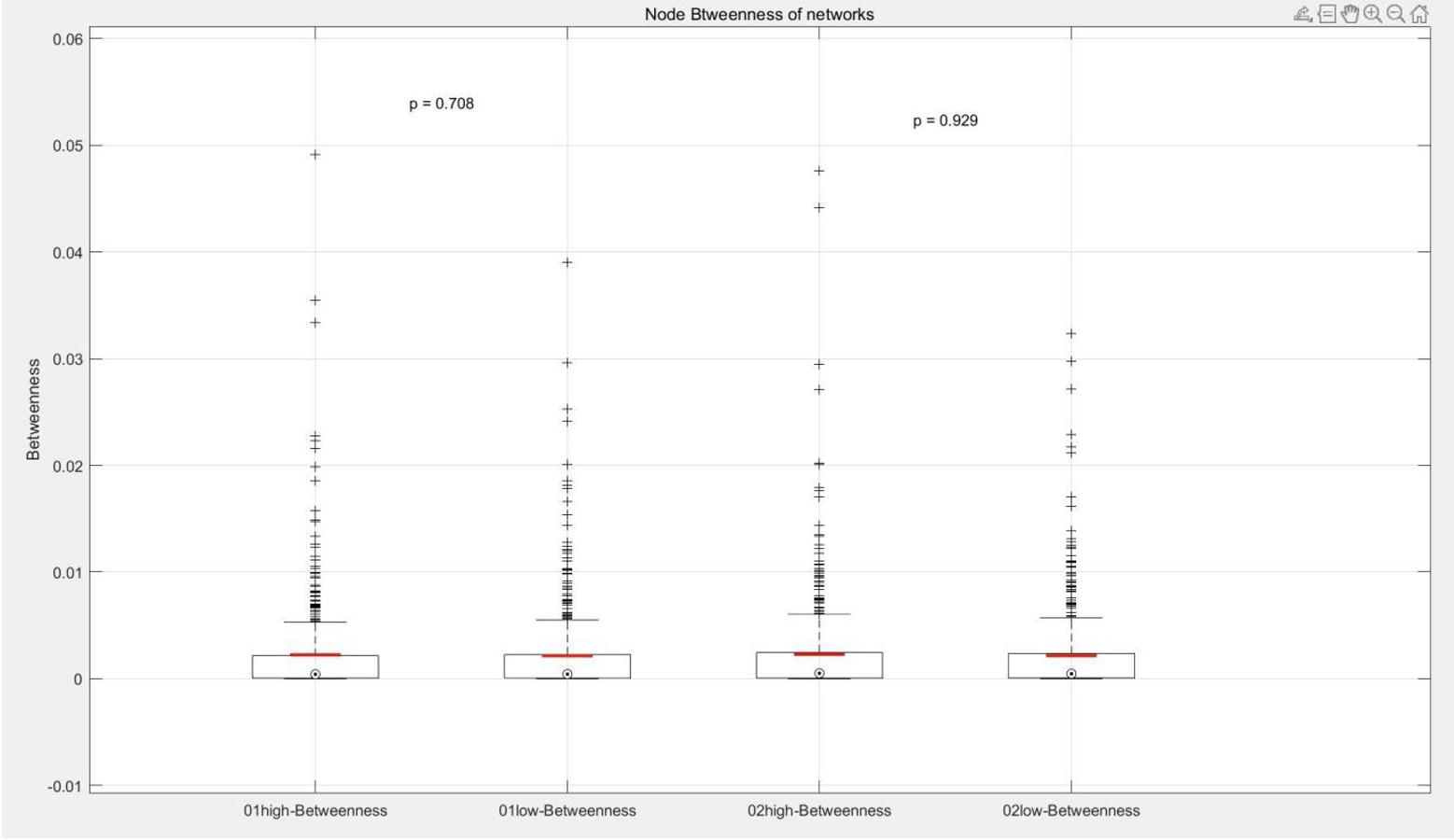}
    \caption{Box plots of Betweenness centrality}
    \label{fig:15}
\end{figure}

As shown in Figure~\ref{fig:15}, the medians of all groups are similar, and while outliers are present, they do not influence the overall comparison between the groups. These outliers likely correspond to specific stations with particularly significant or unusual transit roles within the network. Although the Betweenness centrality of such stations may be critical to their respective networks, their influence does not result in significant differences in the overall distribution of Betweenness centrality.

This analysis suggests that while certain stations may have unique transit roles during specific periods, these roles do not create substantial distinctions between high- and low-frequency networks from a network-wide perspective. This consistency implies that the transit functions and network structure of the stations remain stable, even under varying passenger flows. Furthermore, it highlights the robustness of the network design, where stations can effectively maintain their connection and transfer functions regardless of fluctuations in passenger activity.

\begin{figure}[h]
    \centering
    \includegraphics[width=10cm, height=8cm]{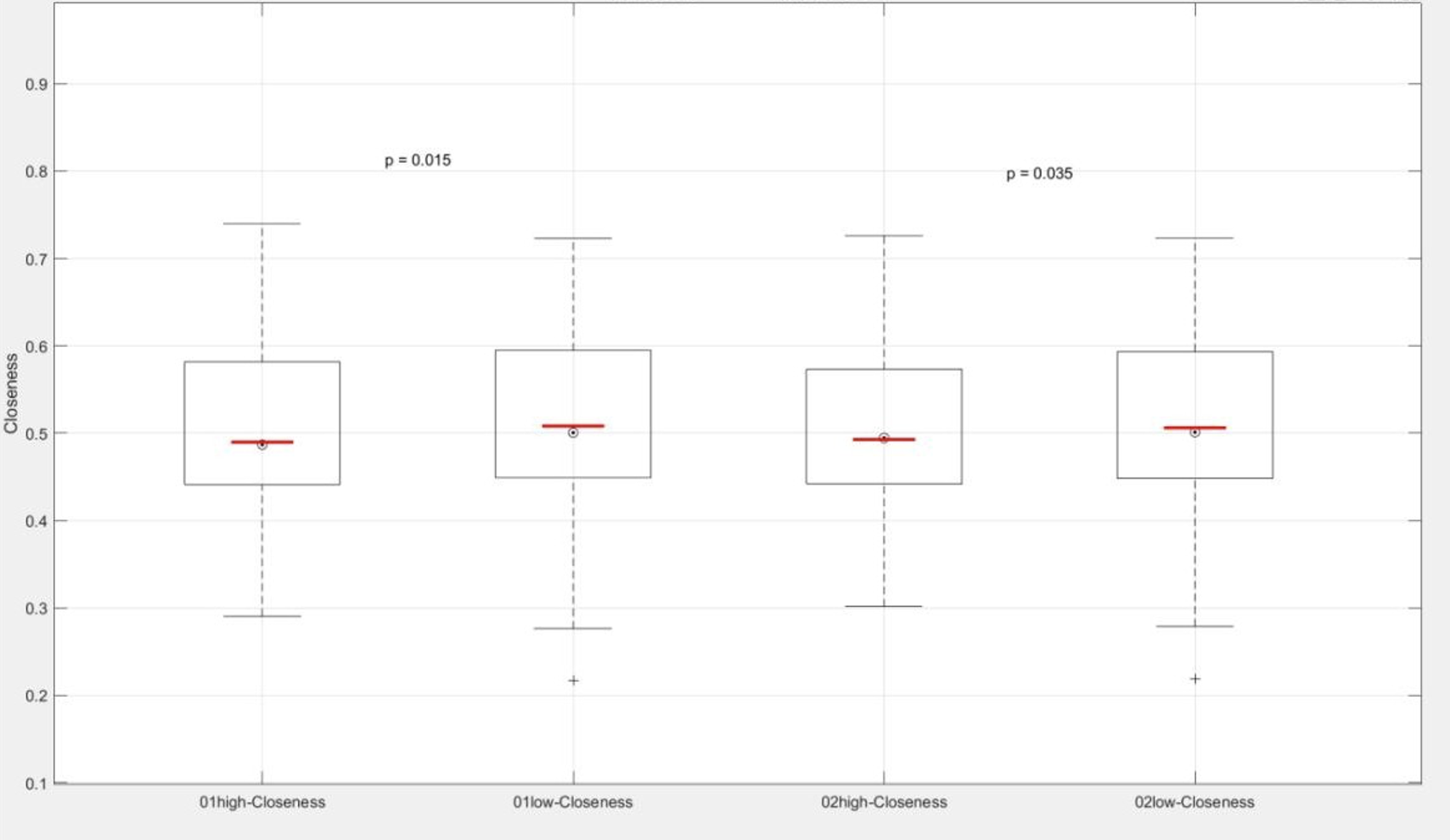}
    \caption{Box plots of closeness centrality}
    \label{fig:16}
\end{figure}

As shown in Figure~\ref{fig:16}, groups 01 and 02 demonstrate a statistically significant difference in closeness centrality. The p-value for group 01 is 0.015, and for group 02 it is 0.035, both of which are below the conventional significance threshold of 0.05. This indicates substantial differences in the closeness centrality of nodes between the two time periods.

Analysis of these differences reveals that the closeness centrality of sites in the low-frequency network is generally higher compared to the high-frequency network. This likely reflects the strategic positioning of sites within the low-frequency network, enabling passengers to reach other stations more quickly. However, given the smaller number of high-frequency passengers, the difference in closeness centrality between the two networks is not pronounced. This suggests that stations in the high-frequency network also provide efficient connectivity, particularly for frequent travelers, such as commuters, who may prioritize time savings and rely on these stations.

Additionally, the upper and lower bounds of closeness centrality in the high-frequency network are higher than those in the low-frequency network. This finding indicates that it is easier to reach other stations within the high-frequency network, further emphasizing its efficiency in catering to passengers with frequent travel needs.

In the low-frequency network, the lower limit of closeness centrality is lower, while the average and median are higher. This suggests that the low-frequency network includes stations located in more remote areas or slightly away from the main routes of the urban transportation network.

Comprehensive analysis of the three node indicators reveals key insights. The significant differences in node degree indicate that some stations bear higher traffic in the low-frequency network, playing a critical role in passengers’ daily travel. Conversely, in the high-frequency network, traffic is more evenly distributed across stations, with passenger flow less concentrated at specific sites. Although the geographical positions of stations remain constant, their importance within the network shifts with passenger traffic patterns.

The analysis of Betweenness centrality shows that the strategic role of stations as passenger flow transit points remains consistent between high- and low-frequency networks. This consistency suggests that, despite variations in overall traffic volume, passenger flow paths and routes between stations are relatively stable.

Differences in closeness centrality highlight the higher accessibility of sites in the high-frequency network, reflecting improved connectivity and efficiency under conditions of high passenger traffic. This allows passengers, particularly frequent travelers, to reach destinations more quickly.

For improving the high-frequency network, given the smaller number of passengers but higher closeness centrality at some nodes, priority should be given to increasing the transportation capacity at these central stations or enhancing their transfer efficiency. Additionally, to address the concentration of high-degree nodes, constructing new routes or increasing the capacity of existing ones could help distribute traffic more evenly and enhance the network’s overall robustness and capacity.

In the low-frequency network, for nodes with lower traffic, increasing direct connections to high-demand nodes could improve their accessibility and appeal. This would enhance the utilization and efficiency of these underused parts of the network, contributing to a more balanced and effective transportation system.

\subsection{Robustness test}

According to the summary of the indicators in the above section, the top 10 nodes with the highest comprehensive scores in the network are selected as the central nodes. The comprehensive score is calculated by summing the Z-score normalized values of node degree, Betweenness centrality, and closeness centrality for each node. This selection criteria ensures a holistic evaluation of the nodes' performance across these three metrics, identifying the most prominent and strategically significant nodes in the network.

The Z-score (Z value) is a standard statistic used to measure the deviation of a data point from the sample mean, expressed in units of the sample standard deviation. Z-score normalization ensures that all indicators are on the same scale, allowing for a fair comparison and aggregation of metrics. The formula for the Z-score is:

\begin{equation}
Z = \frac{(X - \mu)}{\sigma}
\tag{2}
\end{equation}

Where $X$ represents the value of the original data point, $\mu$ is the mean of the sample, and $\sigma$ is the standard deviation of the sample. The $Z$-score calculations standardize the data, eliminating dimensional differences between variables. This standardization facilitates comparisons and enables a comprehensive assessment of the variables, ensuring each metric contributes equally to the overall score.

Using this method, the central nodes of the network were identified, and the results are summarized in Table~\ref{tab:20}. This table highlights the nodes with the highest comprehensive scores, reflecting their significant roles within the network based on their node degree, betweenness centrality, and closeness centrality.

\begin{table}[H]
\centering
\small
\caption{Center Nodes of the High and Low Frequency Network}
\label{tab:20}
\begin{tabularx}{\textwidth}{>{\centering\arraybackslash}X 
                           >{\centering\arraybackslash}X 
                           >{\centering\arraybackslash}X 
                           >{\centering\arraybackslash}X 
                           >{\centering\arraybackslash}X}
\toprule
\textbf{Network} & \textbf{01high} & \textbf{01low} & \textbf{02high} & \textbf{02low} \\
\midrule
\textbf{Central\_Node\_1} & 234 & 234 & 121 & 234 \\
\textbf{Central\_Node\_2} & 121 & 121 & 234 & 121 \\
\textbf{Central\_Node\_3} & 245 & 470 & 431 & 494 \\
\textbf{Central\_Node\_4} & 431 & 494 & 470 & 292 \\
\textbf{Central\_Node\_5} & 470 & 245 & 42 & 245 \\
\textbf{Central\_Node\_6} & 292 & 392 & 245 & 392 \\
\textbf{Central\_Node\_7} & 392 & 431 & 402 & 431 \\
\textbf{Central\_Node\_8} & 154 & 292 & 392 & 42 \\
\textbf{Central\_Node\_9} & 42 & 42 & 292 & 470 \\
\textbf{Central\_Node\_10} & 402 & 357 & 511 & 154 \\
\bottomrule
\end{tabularx}
\end{table}

To explore the robustness of high-frequency and low-frequency passenger traffic networks, this study employs four key indicators: global clustering coefficient, strongly connected components, average path length, and network efficiency. Leveraging optimization techniques, including neural network-based modeling for decision-making, maintenance strategies in networked systems, and adaptive learning algorithms for decentralized architectures, has been shown to enhance network resilience and efficiency \cite{b33, b34, b35}. By identifying central nodes, as described before, a comparative analysis is conducted by examining the changes in these indicators before and after the removal of central nodes. This approach reveals the robustness of high- and low-frequency passenger complex networks when critical nodes fail, providing insights into the networks' ability to maintain connectivity and functionality under such disruptions. 
Tables~\ref{tab:21} and ~\ref{tab:22} present the network features before and after the removal of central sites.

\begin{table}[H]
\centering
\small
\caption{Network Features Before the Removal of Central Sites}
\label{tab:21}
\begin{tabularx}{\textwidth}{>{\centering\arraybackslash}X 
                           >{\centering\arraybackslash}X 
                           >{\centering\arraybackslash}X 
                           >{\centering\arraybackslash}X 
                           >{\centering\arraybackslash}X}
\toprule
\textbf{Complex Network} & \textbf{Global Agglomeration Coefficient} & \textbf{Strong Connectivity Component} & \textbf{Average Path Length} & \textbf{Network Efficiency} \\
\midrule
01high & 0.000963703 & 18 & 2.00754E-06 & 40195.4662 \\
02high & 0.000737876 & 14 & 1.01798E-06 & 45516.9691 \\
01low  & 0.000941457 & 13 & 1.41143E-06 & 38284.88213 \\
02low  & 0.000958783 & 15 & 1.57739E-06 & 41033.7702 \\
\bottomrule
\end{tabularx}
\end{table}

\begin{table}[H]
\centering
\small
\caption{Network Features Before the Removal of Central Sites}
\label{tab:22}
\begin{tabularx}{\textwidth}{>{\centering\arraybackslash}X 
                           >{\centering\arraybackslash}X 
                           >{\centering\arraybackslash}X 
                           >{\centering\arraybackslash}X 
                           >{\centering\arraybackslash}X}
\toprule
\textbf{Complex Network} & \textbf{Global Agglomeration Coefficient} & \textbf{Strong Connectivity Component} & \textbf{Average Path Length} & \textbf{Network Efficiency} \\
\midrule
01high & 0.001157316 & 18 & 1.7097E-06  & 38952.73212 \\
02high & 0.000778501 & 14 & 9.1119E-07  & 43540.0276 \\
01low  & 0.000870484 & 11 & 8.81033E-07 & 46680.90104 \\
02low  & 0.00089554  & 12 & 1.36916E-06 & 40090.89339 \\
\bottomrule
\end{tabularx}
\end{table}

For the high-frequency networks, a slight decrease in the global clustering coefficient was observed after the removal of central nodes. For instance, in the first week, the clustering coefficient dropped from 0.001157316 to 0.000963703. This change indicates that the overall network structure remained relatively stable, though the removal of central nodes slightly affected local connectivity. However, network efficiency showed a significant decrease in both high-frequency networks, underscoring the importance of central nodes in maintaining overall communication efficiency.

Low-frequency networks exhibited different robustness characteristics. Interestingly, the clustering coefficient increased slightly after the removal of central nodes, such as in the 01low network, where it rose from 0.000870484 to 0.000941457. This may reflect better connectivity in other parts of the network or the presence of more alternative paths in low-frequency networks. Nonetheless, similar to high-frequency networks, the average path length in low-frequency networks increased, and network efficiency decreased, emphasizing that central nodes are also critical for the performance of low-frequency networks.

Overall, both high- and low-frequency networks showed structural changes after the removal of central nodes. The decline in the clustering coefficient and the significant reduction in network efficiency for high-frequency networks highlight their dependency on central nodes. The efficiency drop of 5\% and the clustering coefficient reduction of 0.1\% may be attributed to the concentrated traffic flow and limited route diversity in high-frequency passenger movements. Conversely, the increased clustering coefficient in low-frequency networks suggests that other parts of the network retain structural integrity, even with central node removal. However, the increase in average path length and the decrease in network efficiency across both network types reinforce the critical role of central nodes in ensuring efficient network operation.

\subsection{Basic property analysis of complex networks}
\subsubsection{Cluster coefficient}

The agglomeration coefficient measures the extent to which a site's adjacent nodes are interconnected. Specifically, it represents the likelihood that if a site A has direct connections with both sites B and C, there is also a direct connection between B and C. This metric reflects the clustering tendency in the network, capturing the local tightness and formation of small group structures.

The global agglomeration coefficient, in contrast, measures the actual proportion of all possible triangular relationships across the entire network. It provides an overall assessment of the network's clustering tendency, extending the concept of the local agglomeration coefficient to the entire system. This metric reveals how closely nodes are interconnected on a network-wide scale.

In this study, the weighted global agglomeration coefficient is calculated using nx.average\_clustering(G, weight='traffic'), which accounts for the traffic flow between nodes as a weight. High values of the global agglomeration coefficient indicate the presence of many closely connected groups of nodes, suggesting an effective network design capable of dispersing and managing high traffic efficiently.

For Table~\ref{tab:23}, the observed global agglomeration coefficients for high-frequency networks are 0.001157316 and 0.000778501, showing higher values compared to the low-frequency networks, which are 0.000870484 and 0.00089554, respectively. This suggests that high-frequency networks exhibit more closed loops and three-node connections among their nodes. The characteristics of this network structure may be influenced by the daily commuting behavior of high-frequency passengers, who tend to utilize various routes to meet their diverse travel needs.

\begin{table}[H]
\centering
\small
\caption{Global Agglomeration Coefficient of Each Network}
\label{tab:23}
\begin{tabularx}{\textwidth}{>{\centering\arraybackslash}X 
                           >{\centering\arraybackslash}X}
\toprule
\textbf{Complex Networks} & \textbf{Global Agglomeration Coefficient} \\
\midrule
01high & 0.001157316 \\
02high & 0.000778501 \\
01low  & 0.000870484 \\
02low  & 0.00089554 \\
\bottomrule
\end{tabularx}
\end{table}

In contrast, the low-frequency networks display lower and relatively consistent agglomeration coefficients, indicating fewer direct connections between nodes. This simpler and more concentrated network structure likely reflects a tendency among low-frequency passengers to use major traffic routes or direct paths to reach primary destinations, rather than relying on complex networks requiring multipath selections as seen with high-frequency passengers.

The higher agglomeration coefficient observed in high-frequency networks highlights their complex and highly interconnected structure. This complexity is likely due to daily commuters needing to access multiple destinations within the city, such as workplaces, schools, or shopping centers, resulting in more extensive and variable route choices. This structural difference underscores the varying network design requirements for catering to the distinct travel behaviors of high- and low-frequency passengers.
\subsubsection{Strongly connected component}

By analyzing strongly connected components, this study identifies potential weak links or performance bottlenecks, enabling the proposal of improvement measures to optimize network design, enhance risk management, and improve service capability.

On a technical level, the identification of strongly connected components is achieved using a graph traversal algorithm. Specifically, the number of strongly connected components in this study is determined by the strongly\_connected\_components function provided by the network analysis library. This function employs depth-first search (DFS) to systematically explore the network. Each time an unvisited node is encountered during the DFS traversal, a new strongly connected component is identified and explored. This process continues until all nodes in the graph have been fully analyzed, providing a comprehensive understanding of the network's connectivity and structural integrity.

A strongly connected component analysis of high- and low-frequency passenger traffic networks reveals variations in network connectivity across different periods and their behavioral implications. Based on the results in Table~\ref{tab:24}, several conclusions can be drawn.

\begin{table}[H]
\centering
\small
\caption{Strong Connected Component of Each Network}
\label{tab:24}
\begin{tabularx}{\textwidth}{>{\centering\arraybackslash}X 
                           >{\centering\arraybackslash}X}
\toprule
\textbf{Complex Networks} & \textbf{Strong Connectivity Component} \\
\midrule
01high & 18 \\
02high & 14 \\
01low  & 11 \\
02low  & 12 \\
\bottomrule
\end{tabularx}
\end{table}

The comparison of strongly connected components between high- and low-frequency networks highlights notable differences. In the first week, the high-frequency network had significantly more strongly connected components than the low-frequency network (11 vs. 7). In the second week, this difference was smaller but still present (14 vs. 12). These distinctions likely reflect differences in passenger travel behavior. High-frequency networks primarily represent the travel patterns of daily commuters, characterized by more dispersed paths and nodes in the urban transportation network. In contrast, low-frequency networks correspond to non-daily or traveler behaviors, with relatively concentrated travel paths along primary or specific routes.

The higher number of strongly connected components in the high-frequency network suggests that commuters utilize a wider range of paths and connections, contributing to a more dispersed network structure. Conversely, the smaller number of strongly connected components in the low-frequency network indicates a tendency for travelers to use concentrated and predictable routes. These findings align with previous analyses, further emphasizing the differing network dynamics driven by the behaviors of high- and low-frequency passenger groups.

\subsection{Average path length}
The average shortest path length represents the average "cost" of transferring information between nodes or completing a journey within a directed graph. In the context of high- and low-frequency passenger complex networks studied in this paper, this metric provides insights into the average direct accessibility of passengers within the network.

The method employed in this study calculates the average shortest path length for all node pairs within the largest strongly connected component of the directed graph. This calculation is performed using the average\_shortest\_path\_length function from the NetworkX library. The process leverages the depth-first search (DFS) algorithm for weighted graphs, incorporating "normalized traffic" as edge weights. Here, the weight of each edge reflects the relative traffic flow between two sites.

By using these weights in the calculation, the shortest path length is measured based on the cumulative traffic values along the path. Consequently, the resulting average shortest path length reflects not only the spatial characteristics of the network but also the average flow efficiency of passengers within it, providing a comprehensive view of the network's performance. The formula is expressed as follows:

\begin{equation}
Z = \frac{1}{|S|(|S| - 1)} \sum_{\substack{u,v \in S \\ u \neq v}} d(u, v)
\tag{3}
\end{equation}

Where,$|S|$ represents the number of nodes in the largest strongly connected component being analyzed, and $d(u,v)$ denotes the shortest path length from node $u$ to node $v$. The path length is calculated using the weights assigned to the edges, where the weights correspond to standardized traffic values. This means that the shortest path length is a weighted measure reflecting the relative traffic flow between nodes, rather than the actual physical distance or travel time. This approach provides a more nuanced understanding of the network's accessibility and flow efficiency.

From the data in Table~\ref{tab:25}, the high-frequency network (01high and 02high) and the low-frequency network (01low and 02low) exhibit notable differences in average shortest path length. Specifically, the high-frequency network, particularly 01high, has a longer average shortest path length. This suggests that the travel paths of daily commuters are more complex compared to low-frequency travelers.

\begin{table}[H]
\centering
\small
\caption{The Average Shortest Path of Each Network}
\label{tab:25}
\begin{tabularx}{\textwidth}{>{\centering\arraybackslash}X 
                           >{\centering\arraybackslash}X}
\toprule
\textbf{Complex Networks} & \textbf{Average Shortest Path} \\
\midrule
01high & 1.7097E-06 \\
02high & 9.1119E-07 \\
01low  & 8.81033E-07 \\
02low  & 1.36916E-06 \\
\bottomrule
\end{tabularx}
\end{table}

In high-frequency networks, commuters often require more transfers and links, reflecting the need to navigate between various locations such as workplaces, residences, and urban services. This complexity may arise from the larger distances between key destinations or the network's design, which aims to offer broader coverage and more transfer options to accommodate diverse commuting needs.

However, a longer average shortest path length may also highlight potential inefficiencies in the network. For commuters, extended paths can result in increased travel time and higher costs, potentially influencing their travel behavior and even their choices of residence and workplace. This underscores the importance of optimizing network design to balance accessibility with efficiency for frequent travelers.

As observed from Tables~\ref{tab:26}, the low-frequency network in the first week exhibits the highest network efficiency, indicating that travelers in this network generally experience lower "traffic costs" to reach their destinations. This aligns with the shorter average shortest path length observed in the 01low network, suggesting that low-frequency networks provide more direct and efficient travel routes for their users.

\begin{table}[H]
\centering
\small
\caption{Network Efficiencies of Each Network}
\label{tab:26}
\begin{tabularx}{\textwidth}{>{\centering\arraybackslash}X 
                           >{\centering\arraybackslash}X}
\toprule
\textbf{Complex Networks} & \textbf{Network Efficiencies} \\
\midrule
01high & 38952.73212 \\
02high & 43540.0276 \\
01low  & 46680.90104 \\
02low  & 40090.89339 \\
\bottomrule
\end{tabularx}
\end{table}

In contrast, high-frequency networks, particularly in the first week, demonstrate relatively lower efficiency. This reflects the higher flow costs associated with daily commuters who often traverse multiple nodes and connections to complete their journeys. The high-frequency networks serve a broader range of areas to accommodate passengers traveling between work, educational institutions, and other urban locations, which are frequently farther apart. Consequently, while high-frequency passenger routes are more complex and involve more transfers, this complexity also highlights the transportation network's capacity to serve a diverse range of destinations.

The lower efficiency in high-frequency networks suggests potential opportunities for network optimization, especially during peak hours when commuter demand is highest. Reduced path efficiency during these times can exacerbate congestion and delays, increasing travel costs and potentially affecting commuters' travel choices and overall quality of life.

Additionally, the global agglomeration coefficient of the high-frequency network (0.0009679085) is slightly higher than that of the low-frequency network (0.0008830120), and the high-frequency network also features a greater number of strongly connected components. These characteristics indicate that the high-frequency network structure is relatively loose, with weaker site connections, while the low-frequency network exhibits a more compact structure with stronger inter-site connections. This disparity can be attributed to the more complex and diverse travel patterns in high-frequency networks, which include more sites and routes, resulting in a more intricate network topology. The increased indirect connectivity and redundant paths in high-frequency networks lead to a 16\% increase in average path length and a corresponding 5\% decrease in network efficiency.

The Louvain Community Detection Algorithm is a hierarchical clustering approach based on modularity optimization, designed to uncover highly modular community structures within a network. Modularity measures the quality of network segmentation, where a highly modular segmentation contains numerous internal edges (connections within a community) and fewer external edges (connections between communities). 

This information allows traffic planners to allocate resources more effectively, such as enhancing transport capacity in busier communities or introducing customized services in low-frequency areas. By reducing redundant connections between communities and strengthening services within each community, transportation systems can improve efficiency, minimize congestion, and enhance passenger satisfaction. Additionally, community detection provides a foundation for planning new routes and services. Identified communities that diverge from the existing transportation network may indicate opportunities for potential new lines or services tailored to meet emerging passenger demands.

\begin{figure}[h]
    \centering
    \begin{minipage}{0.48\textwidth}
        \centering
        \includegraphics[width=\linewidth, height=7cm]{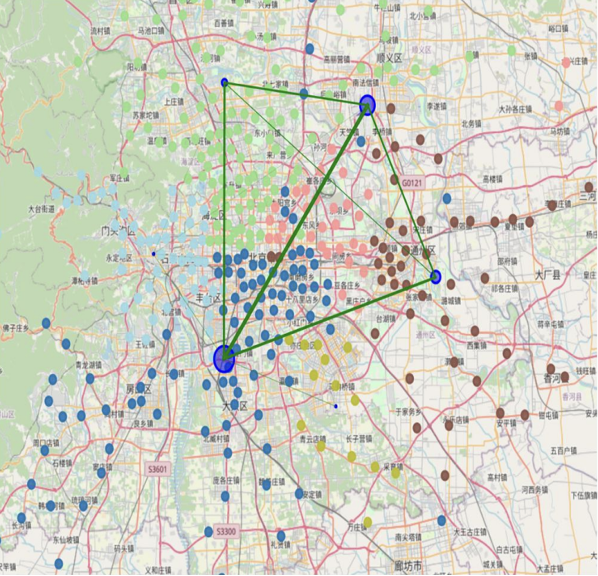}
        \label{fig:171}
    \end{minipage} \hfill
    \begin{minipage}{0.48\textwidth}
        \centering
        \includegraphics[width=\linewidth, height=7cm]{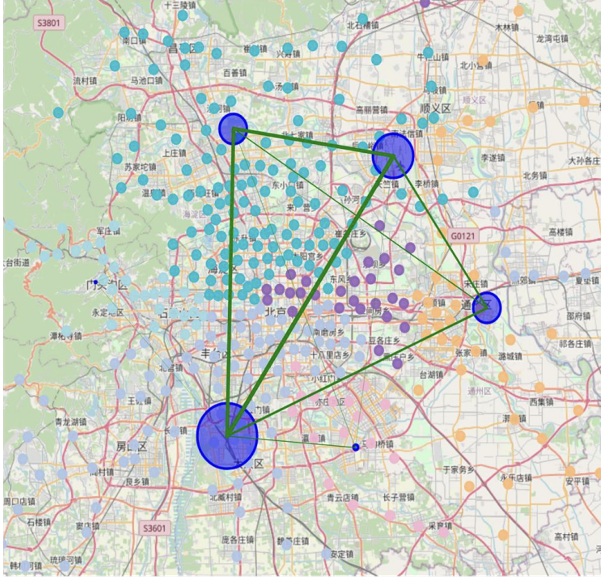}
        \label{fig:172}
    \end{minipage} 
    \caption{High frequency (left) and Low Frequency (right) network community detection in week 1.}
    \label{fig:17}
\end{figure}

\begin{figure}[h]
    \centering
    \begin{minipage}{0.48\textwidth}
        \centering
        \includegraphics[width=\linewidth, height=7cm]{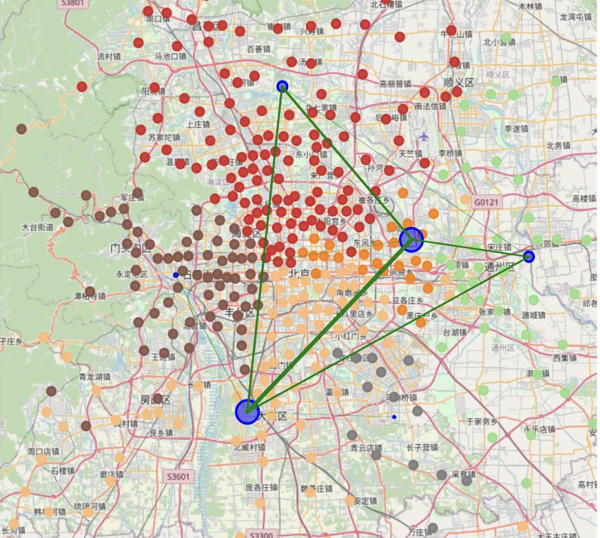}
        \label{fig:181}
    \end{minipage} \hfill
    \begin{minipage}{0.48\textwidth}
        \centering
        \includegraphics[width=\linewidth, height=7cm]{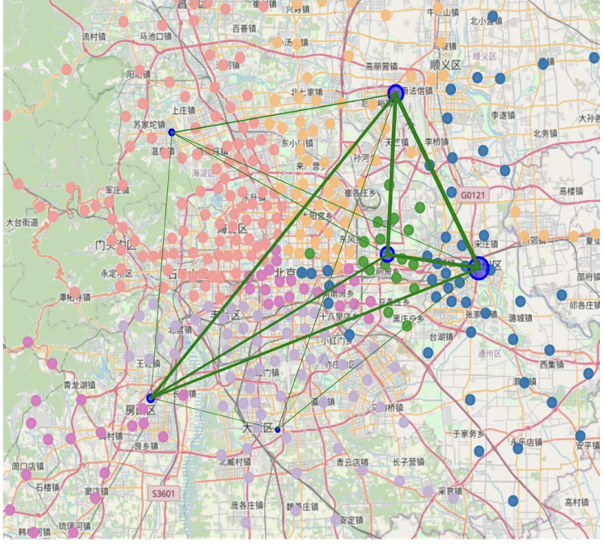}
        \label{fig:182}
    \end{minipage}
    \caption{High frequency (left) and Low Frequency (right) network community detection in week 2.}
    \label{fig:18}
\end{figure}

After performing visual community detection, over 500 cluster sites were grouped into multiple communities, each represented by a distinct color. The flow between communities is visualized using green lines connecting the center points of the communities, with the thickness of the lines corresponding to the flow size. Similarly, the flow within each community is depicted by the size of the circle at the community's center point. These visualizations effectively illustrate the distribution of inter-community and intra-community flows. The results of this analysis are presented in Figures~\ref{fig:17} and Figure~\ref{fig:18}

In the visualization map of the high- and low-frequency passenger networks during the first week, the distribution and connections between central nodes appear similar. However, in the second week, the high-frequency passenger network exhibits larger communities with more stations within each community, whereas the low-frequency passenger network has a greater number of communities with more distinct divisions and closer internal connections.

To further analyze the complex networks following community detection, three indicators are used to characterize the networks. The size represents the number of nodes within a community, offering insight into the scale of individual communities. The average degree is the mean number of edges per node in the network, providing a macroscopic measure of network connectivity. A high average degree indicates that nodes are typically connected to many other nodes, suggesting robust propagation or efficient information exchange within the network.

Modularity, a key metric for evaluating the quality of network community structures, measures the degree of node clustering within communities. High modularity indicates a pronounced community structure, where nodes are more likely to connect with others in the same community rather than external ones. In urban traffic networks, areas with high modularity often correspond to regions with concentrated internal traffic flows, which can serve as indicators of high-traffic zones or regions requiring targeted transportation strategies. The modularity in a network is usually defined by the following formula proposed by Newman \cite{b33}:                                                 
\begin{equation}
L = \frac{1}{2m} \sum_{ij} \left[ A_{ij} - \frac{k_i k_j}{2m} \right] \sigma(c_i, c_j)
\tag{4}
\end{equation}

Where, $A_{ij}$ represents the adjacency matrix of the network, $k_i$, $k_j$ are the degree of nodes $i$ and $j$, $m$ is the total number of edges in the network, and $\sigma(c_i, c_j)$ is an indicator function to detect whether nodes $i$ and $j$ belong to the same community.

The network community metrics presented in Table~\ref{tab:27}, derived from the community structures identified using the Louvain algorithm, highlight the characteristic differences between the high- and low-frequency passenger networks during the first and second weeks. These metrics offer valuable insights into the behavioral patterns of high- and low-frequency passengers, reflecting how their travel preferences and interactions shape the community structures within the urban transportation network.

\begin{table}[H]
\centering
\small
\caption{Each Network Community Indicator}
\label{tab:27}
\begin{tabularx}{\textwidth}{>{\centering\arraybackslash}X 
                           >{\centering\arraybackslash}X 
                           >{\centering\arraybackslash}X 
                           >{\centering\arraybackslash}X 
                           >{\centering\arraybackslash}X}
\toprule
\textbf{Network} & \textbf{Community} & \textbf{Size} & \textbf{Average Degree} & \textbf{Modularity} \\
\midrule
01high & 1 & 112 & 35.51785714 & 0.176978184 \\
       & 2 & 134 & 41.3880597  & 0.176978184 \\
       & 3 & 48  & 20.33333333 & 0.176978184 \\
       & 4 & 47  & 17.74468085 & 0.176978184 \\
       & 5 & 14  & 7.428571429 & 0.176978184 \\
       & 6 & 57  & 22.21052632 & 0.176978184 \\
\midrule
02high & 1 & 28  & 21.07142857 & 0.224282329 \\
       & 2 & 96  & 33.1875     & 0.224282329 \\
       & 3 & 51  & 14.54901961 & 0.224282329 \\
       & 4 & 157 & 41.41401274 & 0.224282329 \\
       & 5 & 68  & 27.23529412 & 0.224282329 \\
       & 6 & 14  & 7.857142857 & 0.224282329 \\
       & 7 & 2   & 1           & 0.224282329 \\
\midrule
01low  & 1 & 122 & 44.85245902 & 0.210709338 \\
       & 2 & 43  & 18.55813953 & 0.210709338 \\
       & 3 & 46  & 19.34782609 & 0.210709338 \\
       & 4 & 2   & 1           & 0.210709338 \\
       & 5 & 19  & 12.21052632 & 0.210709338 \\
       & 6 & 161 & 49.83850932 & 0.210709338 \\
       & 7 & 38  & 15.68421053 & 0.210709338 \\
\midrule
02low  & 1 & 53  & 20.64150943 & 0.224288522 \\
       & 2 & 89  & 30.92134831 & 0.224288522 \\
       & 3 & 17  & 14.35294118 & 0.224288522 \\
       & 4 & 131 & 36.33587786 & 0.224288522 \\
       & 5 & 75  & 24.50666667 & 0.224288522 \\
       & 6 & 50  & 20.56       & 0.224288522 \\
       & 7 & 2   & 1           & 0.224288522 \\
\bottomrule
\end{tabularx}
\end{table}

For high-frequency networks, the larger community sizes likely represent major routes used for daily commuting, while the higher average degree of communities (42) indicates frequent passenger exchanges between stations. This suggests that high-frequency passengers tend to use diverse paths, reflecting a more varied travel pattern. The relatively low modularity (0.40) implies that these communities are not isolated clusters but are situated at critical intersections within the transportation network, allowing for greater flexibility in passenger flow across different communities.

In low-frequency passenger networks, the higher modularity (0.44) indicates more distinct community divisions. This strong internal connectivity may result from the occasional clustering of passengers for specific activities, such as weekend leisure or holiday events, leading to concentrated flows in certain areas. Larger community sizes may correspond to hotspots like tourist attractions or significant service points, which draw centralized traffic.

Urban planners can use these findings to tailor traffic planning strategies to the needs of different passenger groups. For high-frequency passengers, enhancing network connectivity and capacity is crucial, as their travel needs focus on daily, cross-district commuting with diverse paths. Improving network reliability and alleviating congestion in large high-frequency communities will directly enhance travel efficiency. This could involve allocating more public transport resources and increasing service frequency in key areas of high-frequency networks.

For low-frequency passengers, particularly within large communities, ensuring high-quality transportation services is vital to improve their travel experience. Strategies might include optimizing direct services to popular destinations and providing sufficient network capacity to accommodate sudden spikes in demand.

Overall, the mean and modularity indicators in high-frequency networks underscore the importance of inter-community mobility and flexibility in traffic planning, while the same metrics in low-frequency networks highlight the need to focus on destination-centric traffic and intra-community travel demands. This dual approach is critical for designing peak and off-peak transportation strategies that enhance both the efficiency of urban transportation systems and passenger satisfaction.

\subsection{Analysis of Network Characteristics During Peak Hours}
By integrating the cluster site information obtained earlier with the traffic data from the morning and evening peak periods, detailed traffic patterns at the stations are derived. This combined analysis, as shown in Table~\ref{tab:28} and Table~\ref{tab:29}, offers valuable insights into station-level traffic dynamics during peak hours. The results highlight the spatial and temporal distribution of passenger flows across different clustered stations, enabling a deeper understanding of network activity and passenger behavior during critical time periods.

\begin{table}[H]
\centering
\small
\caption{Inter-station traffic after the morning rush hour passenger network in the first week}
\label{tab:28}
\begin{tabularx}{\textwidth}{>{\centering\arraybackslash}X 
                           >{\centering\arraybackslash}X 
                           >{\centering\arraybackslash}X 
                           >{\centering\arraybackslash}X 
                           >{\centering\arraybackslash}X 
                           >{\centering\arraybackslash}X 
                           >{\centering\arraybackslash}X 
                           >{\centering\arraybackslash}X}
\toprule
\textbf{Start-site Node ID} & \textbf{Terminate-site Node ID} & \textbf{Start-site Longitude} & \textbf{Start-site Latitude} & \textbf{Terminate-site Longitude} & \textbf{Terminate-site Latitude} & \textbf{Flow} & \textbf{Standardized Traffic} \\
\midrule
42  & 121 & 116.6307111 & 39.89759435 & 116.453518  & 39.9081348  & 10617 & 0.742170022 \\
42  & 234 & 116.6307111 & 39.89759435 & 116.470691  & 39.91255031 & 8381  & 0.585850112 \\
234 & 121 & 116.470691  & 39.91255031 & 116.453518  & 39.9081348  & 6480  & 0.452950224 \\
154 & 121 & 116.6583586 & 39.88788831 & 116.453518  & 39.9081348  & 5384  & 0.3763283   \\
473 & 134 & 116.5270448 & 39.87229261 & 116.4986155 & 39.89621113 & 5227  & 0.365352349 \\
\bottomrule
\end{tabularx}
\end{table}

\begin{table}[H]
\centering
\small
\caption{Inter-station traffic after the evening rush hour passenger network in the first week}
\label{tab:29}
\begin{tabularx}{\textwidth}{>{\centering\arraybackslash}X 
                           >{\centering\arraybackslash}X 
                           >{\centering\arraybackslash}X 
                           >{\centering\arraybackslash}X 
                           >{\centering\arraybackslash}X 
                           >{\centering\arraybackslash}X 
                           >{\centering\arraybackslash}X 
                           >{\centering\arraybackslash}X}
\toprule
\textbf{Start-site Node ID} & \textbf{Terminate-site Node ID} & \textbf{Start-site Longitude} & \textbf{Start-site Latitude} & \textbf{Terminate-site Longitude} & \textbf{Terminate-site Latitude} & \textbf{Flow} & \textbf{Standardized Traffic} \\
\midrule
121 & 42  & 116.453518  & 39.9081348  & 116.6307111 & 39.89759435 & 8312 & 0.581026286 \\
121 & 234 & 116.453518  & 39.9081348  & 116.470691  & 39.91255031 & 6307 & 0.440855705 \\
234 & 42  & 116.470691  & 39.91255031 & 116.6307111 & 39.89759435 & 5248 & 0.36682047  \\
234 & 154 & 116.470691  & 39.91255031 & 116.6583586 & 39.88788831 & 5219 & 0.29488255  \\
470 & 42  & 116.489658  & 39.91021268 & 116.6307111 & 39.89759435 & 4162 & 0.290897651 \\
\bottomrule
\end{tabularx}
\end{table}

Table~\ref{tab:30} is derived by integrating the basic feature values of each network, providing a comprehensive representation of the overall characteristics of the high- and low-frequency passenger networks throughout the two-week period. This table highlights the distinct structural and operational differences between the two networks, offering valuable insights into their behavior and connectivity patterns.

\begin{table}[H]
\centering
\small
\caption{Basic Feature Values of Each Network}
\label{tab:30}
\begin{tabularx}{\textwidth}{>{\centering\arraybackslash}X 
                           >{\centering\arraybackslash}X 
                           >{\centering\arraybackslash}X 
                           >{\centering\arraybackslash}X}
\toprule
\textbf{Networks} & \textbf{Global Agglomeration Coefficient} & \textbf{Mean Shortest Path} & \textbf{Network Efficiency} \\
\midrule
01high & 0.001157316 & 1.7097E-06  & 38952.73212 \\
02high & 0.000778501 & 9.1119E-07  & 43540.0276  \\
01low  & 0.000870484 & 8.81033E-07 & 46680.90104 \\
02low  & 0.00089554  & 1.36916E-06 & 40090.89339 \\
\bottomrule
\end{tabularx}
\end{table}

Table~\ref{tab:31} and Table~\ref{tab:32} illustrate the basic characteristics of the high- and low-frequency passenger networks, derived using consistent data analysis methods. These tables provide a detailed comparison of the structural and functional features of the networks, offering insights into their connectivity and passenger flow dynamics.

\begin{table}[H]
\centering
\small
\caption{Basic Feature Values of Morning Peak Network}
\label{tab:31}
\begin{tabularx}{\textwidth}{>{\centering\arraybackslash}X 
                           >{\centering\arraybackslash}X 
                           >{\centering\arraybackslash}X 
                           >{\centering\arraybackslash}X}
\toprule
\textbf{Networks} & \textbf{Global Agglomeration Coefficient} & \textbf{Mean Shortest Path} & \textbf{Network Efficiency} \\
\midrule
01high & 1.44388E-07 & 1.41919E-05 & 10179.60573 \\
02high & 1.10448E-07 & 6.93728E-06 & 10827.55082 \\
01low  & 2.26735E-07 & 8.59107E-06 & 10313.71482 \\
02low  & 2.51985E-07 & 1.01485E-05 & 9771.539936 \\
\bottomrule
\end{tabularx}
\end{table}

\begin{table}[H]
\centering
\small
\caption{Basic Feature Values of Evening Peak Network}
\label{tab:32}
\begin{tabularx}{\textwidth}{>{\centering\arraybackslash}X 
                           >{\centering\arraybackslash}X 
                           >{\centering\arraybackslash}X 
                           >{\centering\arraybackslash}X}
\toprule
\textbf{Networks} & \textbf{Global Agglomeration Coefficient} & \textbf{Mean Shortest Path} & \textbf{Network Efficiency} \\
\midrule
01high & 1.44388E-07 & 1.41919E-05 & 10179.60573 \\
02high & 1.10448E-07 & 6.93728E-06 & 10827.55082 \\
01low  & 2.26735E-07 & 8.59107E-06 & 10313.71482 \\
02low  & 2.51985E-07 & 1.01485E-05 & 9771.539936 \\
\bottomrule
\end{tabularx}
\end{table}

Among the network characteristics during peak periods, the average agglomeration coefficient for both high-frequency and low-frequency networks in the morning and evening peaks is significantly reduced compared to the overall period, with a decrease of 99\%, particularly in high-frequency networks. This suggests that during peak hours, the travel network becomes more streamlined, reflecting passengers' need for direct routes, particularly among high-frequency commuters returning home from work.

Regarding the shortest path, the shortest path during peak times increases by approximately 10\%. The high-frequency network exhibits a longer average path length, reflecting more inter-node connections during morning rush hours, while low-frequency passengers experience shorter and simpler travel routes. During the evening peak, however, the average path length becomes relatively short for both networks, indicating less detour and more direct travel paths.

Network efficiency analysis shows a 6\% decrease in network efficiency for both high-frequency and low-frequency networks during the morning peak, while evening peaks exhibit relatively low efficiency for both. This decline likely results from commuters returning home, with complex paths and congestion reducing the overall efficiency of the transportation network.

In summary, peak hours bring notable changes to public transportation networks, including reduced agglomeration coefficients, longer average path lengths, and decreased network efficiency. These observations highlight several issues:

1- Network Congestion: The reduced agglomeration coefficient and increased average path length indicate sparse node connections during peak hours. Passengers may need to pass through more stations to reach their destination, signaling congestion and diminished efficiency due to increased traffic flow.

2- Network Overload: Longer average path lengths and lower efficiency suggest excessive network loads during peak hours. The insufficient capacity of stations and lines to meet peak demand is particularly evident in high-frequency networks.

3- Passenger Detours: Congestion often leads to passengers taking longer, indirect paths to avoid crowded stations or routes, contributing to the longer average path lengths and signaling undercapacity in core transportation areas.

4- Declining Service Levels: Reduced network efficiency indicates bottlenecks and congestion points in the public transportation system. This decline in service levels emphasizes the need for better management and expanded capacity to improve passenger experiences during peak hours.

This study examines the structural characteristics of high- and low-frequency passenger networks and their impact on urban public transport. Key findings include differences in network robustness, connectivity, and efficiency, highlighting distinct travel patterns and network demands.

High-frequency networks show a median node degree of 95 and an average of 158, with a heavy reliance on central nodes. Removing these nodes reduces network efficiency by 5\% and the agglomeration coefficient by 0.1\%. In contrast, low-frequency networks, with a median node degree of 110 and an average of 118, are more robust, maintaining structural integrity even after central node removal.

The global agglomeration coefficient is slightly higher in high-frequency networks (0.0009679085) than in low-frequency ones (0.0008830120). High-frequency networks exhibit more dispersed and complex travel patterns, leading to longer paths (16\% increase) and reduced efficiency (5\% decrease). Low-frequency networks, by comparison, have tighter structures and stronger inter-site connections, reflecting concentrated travel behavior.
\subsection{Section Summary}
Community analysis shows high-frequency networks have lower modularity (0.40) and higher inter-community mobility, while low-frequency networks exhibit higher modularity (0.44), often forming around specific activities or destinations.

Peak periods further emphasize network challenges, with a 99\% drop in the agglomeration coefficient, a 10\% increase in path length, and a 70\% decline in efficiency. High-frequency networks experience greater congestion, with 30\% longer paths and 6\% lower efficiency during evening peaks compared to low-frequency networks.

Recommendations include enhancing central node capacity and decentralizing traffic in high-frequency networks, while improving accessibility and direct connections in low-frequency networks to boost utilization and efficiency.

\section{Conclusion}

Based on the smart card data, this study constructs complex networks of high- and low-frequency travel, analyzing the unique characteristics and behaviors of these groups within the urban public transportation system. Through rigorous data preprocessing, network construction, visualization, and complex network analysis, several significant conclusions have been drawn:

Comprehensive Data Processing and Network Construction: This study conducts an exhaustive cleaning, fusion, and verification of the data to ensure its reliability and quality. By distinguishing high- and low-frequency travelers, the study constructs large-scale networks that provide insight into spatiotemporal travel patterns. High-frequency travelers are predominantly concentrated in urban centers and key hubs, with marked increases during morning and evening peak hours. Low-frequency travelers, on the other hand, exhibit more dispersed travel behaviors.

Network Characteristics and Connectivity: Analysis of node degree, centrality, and agglomeration coefficients reveals distinct roles of high- and low-frequency passengers. The median node degree of high-frequency networks is 95, with an average of 158, while the corresponding figures for low-frequency networks are 110 and 118, respectively. This indicates that stations frequented by low-frequency travelers are more centrally located within the overall network. High-frequency networks demonstrate stronger connectivity but are less robust, with a 0.5\% decrease in network efficiency and a 0.1\% reduction in agglomeration coefficients. Low-frequency networks, while more fragmented, exhibit higher stability and closer connections between nodes.

Community Structure and Mobility: Community detection using the Louvain algorithm highlights differences in modularity and connectivity. High-frequency networks have lower modularity (0.40) but higher average degrees (42), indicating tighter, interconnected, yet flexible community structures. In contrast, low-frequency networks exhibit higher modularity (0.44) and lower average degrees (40), reflecting more distinct communities often formed around specific activities or destinations.

Peak Period Analysis: During peak hours, the agglomeration coefficient decreases by 99\%, the average path length increases by 10\%, and network efficiency drops by 70\%. These changes are indicative of increased congestion and reduced travel efficiency. High-frequency networks, particularly during evening peaks, show a 30\% longer average path length and 6\% lower efficiency compared to low-frequency networks. This reflects the higher complexity and inter-node connections required for commuters during these times.

Policy Recommendations: To address the identified challenges, this study proposes strategies for enhancing transportation networks. For high-frequency networks, increasing service frequency during peak hours, optimizing hub connections, and improving transfer environments are recommended. Enhancing robustness through redundant connections and decentralizing traffic are also critical. For low-frequency networks, improving intra-community connectivity and accessibility to key destinations is essential, particularly during peak demand periods. Tailored strategies, such as increasing shifts or dedicated lines for specific activities, can further improve efficiency.

This research provides a comprehensive understanding of the roles and characteristics of high- and low-frequency travelers within urban transportation systems. Future work could explore the behavioral impacts of varying thresholds for travel frequency classification. For example, different thresholds, such as 25\% or 50\%, may provide additional insights into passenger group behaviors. Furthermore, incorporating weighted importance for stations—such as prioritizing subway hubs over bus stops—could offer a more nuanced analysis of network roles and flow patterns. These advancements would refine the theoretical and practical implications of this study, contributing to the development of more efficient and equitable public transportation systems.

\section*{Acknowledgments}
We extend our appreciation to the research community and transportation authorities for making smart card data and related infrastructure available for academic studies.

\end{document}